\documentclass{article} 
\usepackage{iclr2027_conference,times}

\usepackage{amsmath,amsfonts,bm}

\def\eqref#1{equation~\ref{#1}}

\def\1{\bm{1}}

\DeclareMathAlphabet{\mathsfit}{\encodingdefault}{\sfdefault}{m}{sl}
\SetMathAlphabet{\mathsfit}{bold}{\encodingdefault}{\sfdefault}{bx}{n}

\usepackage{url}
\usepackage{graphicx}
\usepackage{booktabs}
\usepackage{float}

\usepackage[colorlinks,citecolor=gray]{hyperref}

\title{Cross-attention encoding models reveal \\ dynamic spatiotemporal routing across human higher visual cortex}

\author{%
  \normalfont
  \begin{tabular}[t]{@{}l@{\hspace{3em}}l@{}}
    \textbf{Iishaan Inabathini} 
      & \textbf{Margaret M. Henderson} \\
    Carnegie Mellon University 
      & Carnegie Mellon University \\
    \texttt{iishaan@cmu.edu}
      & \texttt{mmhender@cmu.edu} \\[-1.0em]
  \end{tabular}%
}

\iclrfinalcopy 
\begin{document}

\maketitle

\begin{abstract}
Understanding how the brain parses actions and events from time-varying natural inputs is a central challenge in neuroscience. Recent work has used deep neural network (DNN) models to build stimulus-computable fMRI encoding models that predict single-voxel responses to complex natural videos. However, the majority of video-computable encoding models predict responses using simple linear mappings from model tokens, overlooking the spatiotemporal structure shared by video representations and neural responses. Recent cross-attention encoding models address this limitation for static images, enabling flexible stimulus-dependent weighting of image content across space. Here, we extend this framework to naturalistic video, using per-parcel cross-attention to dynamically route features from a self-supervised video model (V-JEPA-2) across both space and time, fitting this model to fMRI responses to short video clips. We compare joint spatiotemporal attention with factorized and selectively constrained alternatives, and find that joint routing improves
predictions of brain responses to held-out videos across higher visual regions, most consistently in lateral and dorsal visual areas associated with dynamic motion perception. Moreover, our method provides interpretable, stimulus-specific attention maps that dynamically follow moving objects, revealing which locations and temporal moments contribute to each neural response.
We further show that attention maps from parcels in different category-selective networks (face-, body-, scene-selective) differentially weight content in accordance with expected semantic selectivity.
Together, this work provides a new computational framework for understanding how visual information is adaptively weighted by cortical populations during dynamic visual perception.

\end{abstract}

\section{Introduction}
Human observers effortlessly parse objects, events, and actions from a dynamic visual input, supported by computations in higher dorsal, ventral, and lateral visual cortex \citep{ungerleider1982two, Pitcher2021, Grill-Spector2014TheCategorization}. 
Recent work using methods such as functional magnetic resonance imaging (fMRI) along with 
deep neural network (DNN)-based computational models, has substantially advanced understanding of how the higher visual system processes static content such as natural scenes \citep{yamins2016goal, doerig2023neuroconnectionist}. However, while a number of recent studies have applied similar methods to video-evoked data \citep[e.g.,][]{sartzetaki2025one, garcia2025modeling, Tang2025}, we still know comparatively much less about the neural basis of dynamic visual perception.

Modeling neural responses to dynamic videos is challenging due to the spatiotemporal complexity of both videos and video-evoked neural responses. Visual cortex populations, as measured with fMRI, integrate information across time nonlinearly \citep{hasson2008hierarchy, zhou2018compressive, Groen2022}, making it challenging to identify which temporal moment is most predictive of the measured neural response. In addition to temporal structure, natural videos are spatially complex, often including multiple objects or agents that compete for neural representation \citep{zoccolan2005multiple, reddy2007category}, which similarly makes it challenging to identify which spatial region of an image is responsible for driving a neural population. 
Evidence suggests that neural populations in the higher visual system respond to a preferred stimulus at multiple possible locations within a large spatial receptive field \citep{rust2010selectivity}, and their response to the stimulus may depend on other objects \citep{zoccolan2005multiple, reddy2007category, bao2018representation} or surrounding context \citep{bar2004visual}. The resulting neural response properties may then be described as having spatial selectivity that is stimulus-dependent, as opposed to being driven by features within a fixed spatial receptive field. In videos, where objects can also move in time, the relative weighting of information across time may similarly depend on the input, which could result in adaptive spatiotemporal weighting that depends on characteristics of the input video.

\begin{figure}[t]
    \centering
    \includegraphics[width=0.90\linewidth]{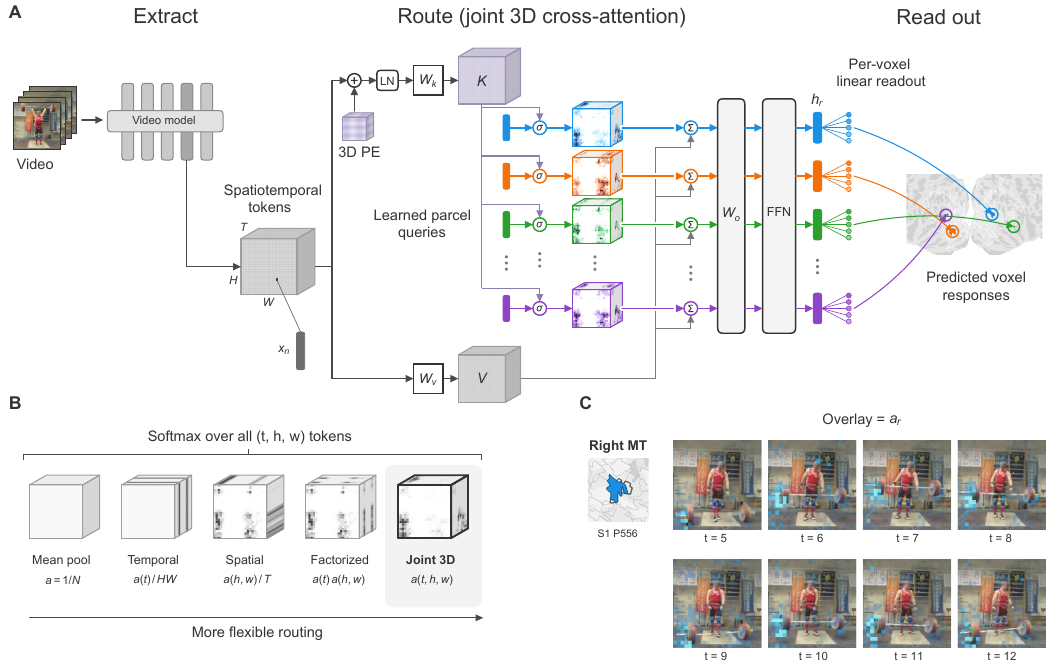}
    \caption{Cross-attention encoding model. \textbf{(A)} A video backbone (V-JEPA~2) encodes each clip into a grid of spatiotemporal tokens. A learned query for each cortical parcel attends over the tokens, and the attended feature predicts the responses of the parcel's voxels through a voxel-specific linear readout. \textbf{(B)} Constrained routing models, shown as example attention patterns over the token grid with the form of their attention weights. \textbf{(C)} Attention of the joint model for a parcel in right MT, overlaid on selected frames of a test clip of a weightlifter.}
    \label{fig:fig1-teaser}
\end{figure}

To model such adaptive stimulus-dependent weighting of visual features by neural populations, recent work has incorporated attention operations, commonly used in Transformer neural networks \citep{vaswani2023attentionneed, dosovitskiy2021imageworth16x16words}, into fMRI encoding models \citep{adeli2026transformerbrainencodersexplain, hwang2025insilico}. While standard fMRI voxelwise encoding model approaches map from DNN features to voxel responses using a simple linear mapping \citep{naselaris2011encoding}, these newer approaches use a cross-attention module with a region-specific query vector to perform a more flexible routing operation that can improve model performance in higher visual regions \citep{adeli2026transformerbrainencodersexplain, hwang2025insilico}. However, it is not clear whether and how such approaches can be extended to modeling brain responses to video stimuli, where spatial and temporal information may be independently and/or jointly reflected in neural response properties. 

Here, we present a new cross-attention encoding model framework for predicting voxelwise fMRI activation in response to natural video clips. Using fMRI data from the BOLD Moments Dataset (BMD; \cite{lahner2024bold}) and features extracted from a self-supervised video model (V-JEPA-2 \citep{assran2025vjepa2selfsupervisedvideo}), we compare the predictive accuracy of models using spatial, temporal, and joint spatiotemporal cross-attention. Our contributions are as follows:

\begin{itemize}
\item We extend previous work by showing that a cross-attention readout leveraging spatial attention over video tokens can improve predictive accuracy of voxelwise encoding models for video-evoked fMRI data, across multiple DNN backbones.
\item We show that the novel addition of joint spatiotemporal attention boosts encoding model accuracy across several higher visual regions, notably in lateral and parietal visual cortex.
\item We demonstrate how joint spatiotemporal attention leads to more interpretable and accurate attention maps compared to factorized attention, tracking the movement of relevant objects across a video clip.
\item We show dissociations between the attention maps derived from different functionally-localized higher visual cortex regions, with attention weights dynamically tracking the location of region-relevant video content.
\end{itemize}
\section{Related Work}

\textbf{Neural coding during dynamic visual perception.}
Classic models of vision suggest that static object recognition and dynamic spatial vision are anatomically segregated into the ventral and dorsal stream, respectively \citep{ungerleider1982two, goodale1992, Grill-Spector2014TheCategorization}, with dorsal areas (MT, MST) key to visual motion processing \citep{albright1984, andersen1997}. Recent work also provides evidence for a lateral visual pathway specializing in dynamic social perception \citep{pitcher2011, Pitcher2021, mcmahon2023}, including regions along the superior temporal sulcus (STS). Other research on temporal dynamics has suggested a hierarchical posterior-to-anterior organization based on temporal receptive window size \citep{hasson2008hierarchy, Groen2022, zhou2018compressive}. Here, we develop a novel strategy to computationally model higher visual cortex during dynamic natural event perception.

\textbf{Category-selective organization in visual cortex.}
Anatomically localized regions of ventral and lateral visual cortex exhibit functional selectivity for human faces (FFA, OFA, STS), human bodies (EBA), and scenes (PPA, TOS, RSC), among other categories \citep{Epstein1998, Kanwisher1997, Downing2006, epstein2019scene}. These regions were classically identified using isolated objects, with recent work extending this using naturalistic fMRI and computational modeling (\cite{Khosla2022, RatanMurty2021, Luo2023a}, and others). However, because natural videos often include multiple categories that interact over time, modeling category-selective responses during natural video perception remains challenging. Here, we show how joint spatiotemporal attention can be used to dissociate the features driving responses in distinct category-selective networks during complex video perception.

\textbf{Voxelwise fMRI encoding models.}
To model cortical fMRI responses during complex natural vision, forward encoding models construct a stimulus-computable function mapping from the input (i.e., a stimulus viewed by an observer) to the response of an fMRI voxel \citep{naselaris2011encoding, Serences2012, Wehbe2014}. This is typically done by extracting stimulus features from a
task-optimized DNN model, then learning a voxel-specific mapping  from the DNN activations to the individual voxel's response \citep{naselaris2011encoding, yamins2016goal}. Benchmarking the cross-validated accuracy of DNN encoding models can be used
to evaluate the alignment of DNN models and neural representations \citep{Schrimpf2020integrative, Wang2023, conwell2024large, sartzetaki2025one, doerig2023neuroconnectionist}.

The voxel-specific mapping is most commonly linear \citep{naselaris2011encoding}, and is often preceded by spatial and/or temporal averaging to handle high-dimensional DNN features.
Other approaches use a
``factorized'' strategy which applies a shared spatial weighting function across feature maps \citep{Klindt2017, StYves2018, lurz2021generalization, Henderson2023b}, assuming separable and input-invariant spatial selectivity.
Recent work replaces this with a cross-attention module that dynamically weights space based on the input \citep{adeli2026transformerbrainencodersexplain, hwang2025insilico}. Here, we extend this flexible routing strategy to the spatiotemporal domain. 

\textbf{Video encoding and decoding models for fMRI.}
Large-scale video fMRI datasets \citep{lahner2024bold, gifford2025algonautsproject2025challenge} have spurred efforts to build predictive video encoding models based on DNN features
\citep{sartzetaki2025one, Tang2025, Hofling2026, gamal2026pixel, han2024investigating, alkarkari2025dynamics, garcia2025modeling, Pushpita2025}. Other work uses video encoding models to perform video reconstruction \citep{chen2023cinematic, lu2024animate, Yeung_2025_CVPR}, or to synthesize an optimal video stimulus \citep{tang2026nevoneuralguidedevolutionaryvideo}. Recent papers find that V-JEPA-2 features \citep{assran2025vjepa2selfsupervisedvideo} achieve top predictive accuracy among video models \citep{Tang2025, tang2026nevoneuralguidedevolutionaryvideo, Hofling2026}.
Our approach is distinct in introducing a spatiotemporal cross-attention framework that yields interpretable maps of how video features are weighted over space and time by cortical populations.

\section{Methods}

\subsection{Video cross-attention encoding model framework}
\label{sec:methods-framework}

We propose a cross-attention encoding model framework for video-evoked fMRI responses (Fig.~\ref{fig:fig1-teaser}) with three goals. First, each cortical population should read out from a video in a stimulus-dependent way, selecting \emph{where} and \emph{when} in the clip to draw features from, rather than relying on a fixed receptive field or on features pooled over the video. Second, the contribution of each form of routing should be testable: spatial, temporal, factorized and joint spatiotemporal routing are constrained versions of the same attention operation, so they can be compared with the backbone, readout and training held fixed. Third, the routing should be interpretable: the attention weights form a stimulus-specific map over the video that can be compared directly with its content. Following cross-attention encoding models for static images \citep{adeli2026transformerbrainencodersexplain, hwang2025insilico}, a learned query for each brain parcel attends over the token grid of a frozen video backbone, and we extend the attention to range jointly over time and space.

Below, we describe the model, the constrained comparison models and the parcel-based queries. Section~\ref{sec:methods-data} describes the data, Section~\ref{sec:methods-fitting} the fitting, evaluation and ridge regression baseline, and Appendix~\ref{app:catsel} the analyses of category selectivity in the attention maps.

\textbf{Spatiotemporal token grid.} A frozen video backbone maps each clip to a grid of tokens $X \in \mathbb{R}^{T \times H \times W \times D}$, with token $x_n \in \mathbb{R}^{D}$ at grid position $(t_n, h_n, w_n)$ and $N = THW$ tokens per clip ($T = 16$, $H = W = 24$, $D = 1408$ and $N = 9216$ for our backbone; Appendix~\ref{app:implementation}). Each channel is z-scored over training clips. The tokens are computed once; all learning takes place in the readout.

\textbf{Cross-attention readout.} Each parcel $r$ has a learned query $q_r \in \mathbb{R}^{D}$ that does not depend on the stimulus. Keys and values are computed from the tokens as
\[
    k_n = W_k\,\mathrm{LN}\big(x_n + \alpha\,\mathrm{PE}(t_n, h_n, w_n)\big), \qquad v_n = W_v\,x_n,
\]
where $W_k, W_v \in \mathbb{R}^{D \times D}$, LN is layer normalization, and PE is a fixed sinusoidal 3D positional encoding whose $D$ channels are split into three blocks encoding $t$, $h$ and $w$ ($\alpha = 1$). Because the positional encoding enters the keys but not the values, a query can prefer locations and moments as well as content, while the values carry content only. The attention of parcel $r$ is a single softmax over all $N$ tokens,
\[
    a_{rn} = \frac{\exp\!\big(q_r^\top k_n / \sqrt{D}\big)}{\sum_{m=1}^{N} \exp\!\big(q_r^\top k_m / \sqrt{D}\big)}, \qquad \sum_{n} a_{rn} = 1,
\]
which, reshaped to the token grid, gives the attention map $a_r(t,h,w)$ (Fig.~\ref{fig:fig1-teaser}). The fixed query encodes a stable preference, and the map shows where and when a clip meets it. The attended feature $z_r = \sum_n a_{rn} v_n$ is passed through an output projection and a residual feed-forward network,
\[
    h_r = W_o z_r, \qquad h_r \leftarrow h_r + \mathrm{FFN}\big(\mathrm{LN}(h_r)\big),
\]
where the FFN is a two-layer MLP with a GELU nonlinearity and hidden width $4D$. We use a single attention head, and $W_k$, $W_v$, $W_o$ and the FFN are shared across parcels, so parcels differ only in their query and voxel readout. Each voxel is predicted from the output $h_r$ of its parcel by a voxel-specific linear readout, trained jointly with the attention module by gradient descent. All voxels in a parcel thus share one attention map but weight the attended features differently.

\textbf{Constrained routing models.} To isolate the contributions of spatial, temporal and coupled spatiotemporal routing, we compare the joint model with models that differ from it only in their attention weights. Their attention factorizes into a temporal and a spatial distribution, $a_r(t,h,w) = a_r^{\mathrm{T}}(t)\, a_r^{\mathrm{S}}(h,w)$, and each factor is uniform, fixed or routed. A uniform factor weights all frames (or locations) equally. A fixed factor is a softmax over learned logits for each parcel, a stimulus-independent receptive field that is the same for every clip. A routed factor is a softmax of the parcel's query against keys computed, with their own key projection and positional encoding, from tokens averaged over the other axis (the spatially averaged token of each frame, or the temporally averaged token at each location). We focus on four of these models:

\begin{itemize}
    \item \textbf{Mean pool} (both factors uniform, $a_{rn} = 1/N$): every parcel reads the same clip-average feature, with no routing.
    \item \textbf{Temporal} ($a_r^{\mathrm{T}}$ routed, $a_r^{\mathrm{S}}$ uniform): the parcel selects moments but pools over space.
    \item \textbf{Spatial} ($a_r^{\mathrm{T}}$ uniform, $a_r^{\mathrm{S}}$ routed): the parcel selects locations but pools over time, as in spatial routing for static images applied to the time-averaged video.
    \item \textbf{Factorized} (both factors routed): the parcel selects moments and locations, but attends to the same locations at every moment, so it cannot follow moving content.
\end{itemize}
Because the joint model's attention is a single softmax over all tokens, a parcel can attend to different locations at different moments. Comparing the joint and factorized models therefore isolates the benefit of coupled spatiotemporal routing, and comparing the spatial and temporal models with mean pooling isolates the benefit of routing along each axis.

\textbf{Parcel-based queries.} We assign one query to each parcel of the Schaefer 1000-parcel, 7-network atlas \citep{schaefer2018local}, resampled to the voxel grid of the dataset (Appendix~\ref{app:data}), which gives 423--433 queries per subject, including one for voxels without a parcel label. We judged this the appropriate granularity: fine enough for differences within and between functional regions to emerge in the attention maps, and coarse enough that each map is estimated from many voxels (Appendix~\ref{app:parcel-queries}).

\subsection{Functional MRI video dataset}
\label{sec:methods-data}

\textbf{BOLD Moments Dataset.} We use the BOLD Moments Dataset (BMD; \cite{lahner2024bold}), in which 10 subjects watched 1,102 naturalistic 3\,s video clips of everyday events during 3T fMRI (2.5\,mm isotropic voxels, TR 1.75\,s): 1,000 training clips shown 3 times each and 102 test clips shown 10 times each. We predict the single-trial response amplitudes released with the dataset (GLMsingle betas; \cite{Prince2022ImprovingGLMsingle}), averaged over the repeats of each clip, in the 17.2--17.5k voxels per subject that lie in cortex with reliable responses to the videos or in the functional ROIs (preprocessing details in Appendix~\ref{app:data}). We also use the per-voxel noise ceiling released with the dataset, computed for responses averaged over the 10 test repeats.

\textbf{Functional ROIs.} We use the 23 ROIs released with the dataset, pooled across hemispheres: retinotopic ROIs from the atlas of \citet{wang2015probabilistic}, MT and parietal ROIs from the HCP multimodal parcellation \citep{glasser2016multi}, and category-selective ROIs defined in each subject with a video-based functional localizer (Appendix~\ref{app:data}).
To summarize prediction accuracy (Fig.~\ref{fig:fig2}), we pool the ROIs into four groups:
early visual (V1--V3, dorsal and ventral, V3ab and hV4), lateral/face/body (MT, EBA, FFA, OFA, LOC, STS), scene (PPA, RSC, TOS) and parietal (IPS0, IPS1-2-3, 7AL, BA2, PFop, PFt). We also report all modeled voxels and the voxels outside all ROIs. The groups follow a clustering of the ROIs by their responses to the videos (Appendix~\ref{app:roi-clustering}).

\subsection{Model fitting and evaluation procedures}
\label{sec:methods-fitting}

We use frozen features from layer 32 of 40 of V-JEPA~2 ViT-g/16 \citep{assran2025vjepa2selfsupervisedvideo}, a self-supervised video model whose features give the most accurate fMRI predictions among current large video models \citep{Tang2025, tang2026nevoneuralguidedevolutionaryvideo, Hofling2026}; layer 32 gave the most accurate predictions in a ridge-regression sweep over the layers of several video and image backbones (Appendix~\ref{app:layer-sweep}). To test whether our results depend on the backbone and its depth, we also fit the joint and all constrained models to features from an early and a late layer of another video model and two image models (Appendix~\ref{app:routing-sweep}). Each subject is fit separately, with the mean squared error over all voxels as the loss. Every model is an ensemble of 10 members, each trained on a different random split of the 1,000 training clips into 900 clips for training and 100 for early stopping, and, following \citet{adeli2026transformerbrainencodersexplain}, the test predictions of the members are averaged voxel by voxel before evaluation. We train with AdamW and early stopping, with the same hyperparameters for all models, backbones and subjects. As a standard linear encoding model, we also fit voxelwise ridge regression to the same backbone layer, averaged over time and space into one 1408-dimensional vector per clip, with the same splits and ensembling (token extraction, optimization and ridge details in Appendix~\ref{app:implementation}). For each voxel, we compute the Pearson correlation $r$ between the ensemble predictions and the measured responses across the 102 test clips and report the signed $r^2 = \mathrm{sign}(r)\,r^2$.

\section{Results}
\label{sec:results-performance}

\begin{figure}[t]
    \centering
    \includegraphics[width=0.75\linewidth]{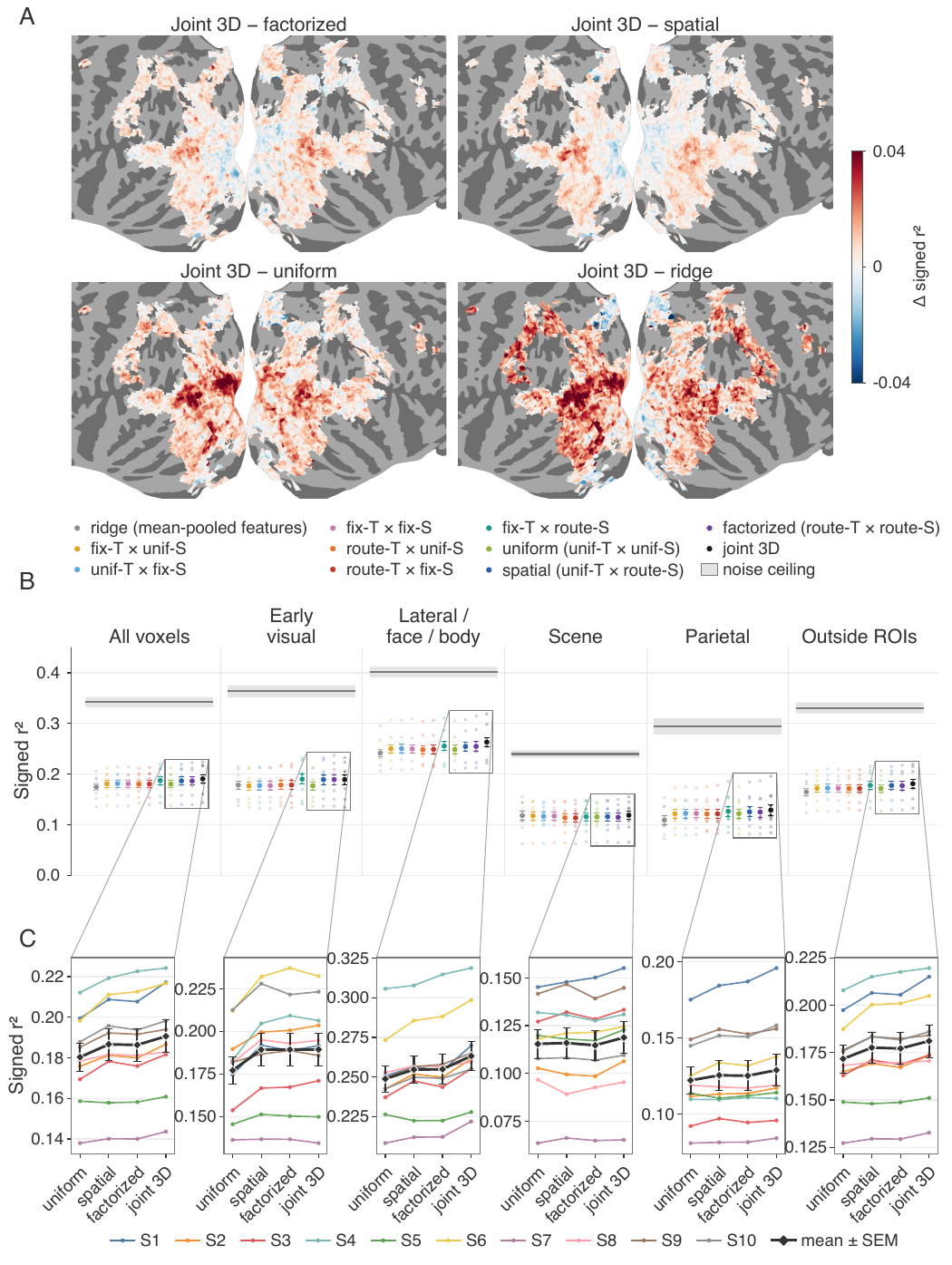}
    \vspace{-0.2in}
    \caption{Comparing accuracy of joint spatiotemporal attention versus other mappings
    (V-JEPA~2 ViT-g, layer 32; 102 test clips). \textbf{(A)} Joint model minus the factorized, spatial and mean-pool (``uniform'') models and ridge regression on mean-pooled features in voxelwise signed $r^2$, averaged over subjects (fsaverage flatmap, unthresholded; each subject in Appendix~\ref{app:flatmaps}). \textbf{(B)} Region-mean signed $r^2$ of every routing configuration (T = temporal factor, S = spatial factor; each uniform, fixed or routed), the joint model and ridge regression. Large dots: mean $\pm$ SEM over subjects; small dots: individual subjects; grey band: noise ceiling (mean $\pm$ SEM). Region means include only voxels with a noise ceiling $\geq$ 5\%. \textbf{(C)} Zoomed inset from (B); shows region means of the mean-pool, spatial, factorized and joint models in each subject (colored lines) and their mean $\pm$ SEM (black).
    }
    \label{fig:fig2}
\end{figure}

\begin{table}[!htbp]
\centering
\footnotesize
\caption{Region-mean signed $r^2$ on the 102 test clips (V-JEPA~2 ViT-g, layer 32; mean over 10 subjects; voxels with noise ceiling $\geq$ 5\%). The best model in each row is in bold. All routing configurations and the other backbones are given in Table~\ref{tab:routing-sweep}.}
\label{tab:region-means}
\begin{tabular}{lrrrrrrr}
\toprule
\textbf{Region} & \textbf{Ridge} & \textbf{Mean pool} & \textbf{Temporal} & \textbf{Spatial} & \textbf{Factorized} & \textbf{Joint} & \textbf{Noise ceiling} \\
\midrule
All voxels & .174 & .180 & .180 & .187 & .186 & \textbf{.191} & .34 \\
Early visual & .178 & .177 & .178 & \textbf{.189} & \textbf{.189} & \textbf{.189} & .36 \\
Lateral/face/body & .242 & .249 & .248 & .255 & .255 & \textbf{.263} & .40 \\
Scene & .118 & .115 & .114 & .116 & .115 & \textbf{.119} & .24 \\
Parietal & .109 & .122 & .121 & .125 & .125 & \textbf{.129} & .29 \\
Outside ROIs & .165 & .172 & .171 & .178 & .177 & \textbf{.181} & .33 \\
\bottomrule
\end{tabular}
\end{table}

\textbf{Comparing model performance across attention types.} We first examine performance averaged across all voxels, then examine region-specific differences. Across all modeled voxels, we observed a substantial advantage for joint spatiotemporal routing, with the joint model having higher $r^2$ than every constrained model and than ridge regression in all 10 subjects (Fig.~\ref{fig:fig2}B, Table~\ref{tab:region-means}), including the closest, the spatial model (0.191 vs.\ 0.187; $t(9) = 5.85$, $p_{\mathrm{FDR}} < 0.001$). Unless stated otherwise, we compare models with two-sided paired $t$-tests of region-mean signed $r^2$ across the 10 subjects (FDR corrected; $p_{\mathrm{FDR}} < 0.05$; \cite{benjamini1995controlling}); see Table~\ref{tab:fig2_ttests}.

The constrained models show which kind of routing produces this gain. Of the 0.010 by which the joint model exceeds mean pooling, 0.006 is recovered by routing over space alone (spatial minus mean pool: $t(9) = 5.07$, $p_{\mathrm{FDR}} < 0.001$; see Table~\ref{tab:spatial-tests} for additional comparisons).
This extends to video previous work showing an advantage of stimulus-dependent spatial routing over linear readouts for static images \citep{adeli2026transformerbrainencodersexplain}. This pattern holds for every backbone and layer we tested, with the exception of scene-selective cortex and for the last stage of CLIP (Appendix~\ref{app:spatial-routing}). Routing over time alone does not improve over uniform pooling (temporal model, 0.180), and adding temporal routing to spatial routing does not improve beyond spatial routing alone (factorized model, 0.186). We also find that the benefit of spatial routing depends on the selection being driven by the stimulus: a fixed spatial distribution learned for each parcel, which acts as a static receptive field, does not exceed the performance of mean pooling (0.181; Table~\ref{tab:routing-sweep}, ablations in Appendix~\ref{app:ablation}). 
These results provide insight into what drives the benefit of the joint spatiotemporal routing mechanism: it allows parcels to attend to different locations at different moments in a stimulus-dependent manner.

This benefit of coupled routing differs between regions. On the flatmaps, the advantage of the joint model over the factorized and spatial models is concentrated in lateral occipitotemporal cortex (Fig.~\ref{fig:fig2}A). At the level of functionally-defined ROIs, the advantage is largest in lateral/face/body regions, where the joint model reaches 0.263 (66\% of the noise ceiling) and exceeds the factorized, spatial, mean-pool and ridge regression models 
in every subject (joint minus factorized: $t(9) = 6.60$, $p_{\mathrm{FDR}} < 0.001$; Fig.~\ref{fig:fig2}C; individual ROIs in Appendix~\ref{app:roi-routing}). 
Parietal regions similarly show an advantage for joint routing over all other models (Table \ref{tab:fig2_ttests}). 
In early visual cortex, by contrast, the spatial, factorized and joint models are tied (0.189; joint minus factorized: $t(9) = -0.03$, $p_{\mathrm{FDR}} = 0.98$), and all three improve on mean pooling by 0.012. Stimulus-dependent selection of spatial locations thus accounts for the whole gain in early visual cortex, whereas lateral and parietal regions
additionally benefit from following content over time. 
In scene-selective regions, we find more mixed evidence for a benefit of spatial or temporal routing. Neither the joint model nor the spatial model significantly exceed the the ridge baseline (Table \ref{tab:fig2_ttests}, Table \ref{tab:spatial-tests}). 
This may reflect the role of scene-selective cortex in encoding global scene layout, which features averaged over the whole clip may sufficiently capture (Appendix~\ref{app:roi-clustering}).

The advantage of coupled routing in lateral regions is not specific to V-JEPA~2 (Appendix~\ref{app:routing-sweep}). With features from the late layers of a second video model (VideoMAE~v2) and two image models (DINOv2 and CLIP), the joint model is again more accurate than the factorized model in lateral/face/body regions (joint minus factorized: $t(9) = 2.88$, $2.49$ and $3.47$; $p_{\mathrm{FDR}} = 0.027$, $0.048$ and $0.011$, respectively; FDR-corrected over the 192 tests of the sweep). The pattern depends on the features, however. With features from early layers, the joint model is less accurate than models that route over space alone, and for the last stage of CLIP, ridge regression on mean-pooled features is the most accurate model in every region except early visual (Table~\ref{tab:routing-sweep}).

\begin{figure}[t]
    \centering
    \includegraphics[width=0.90\linewidth]{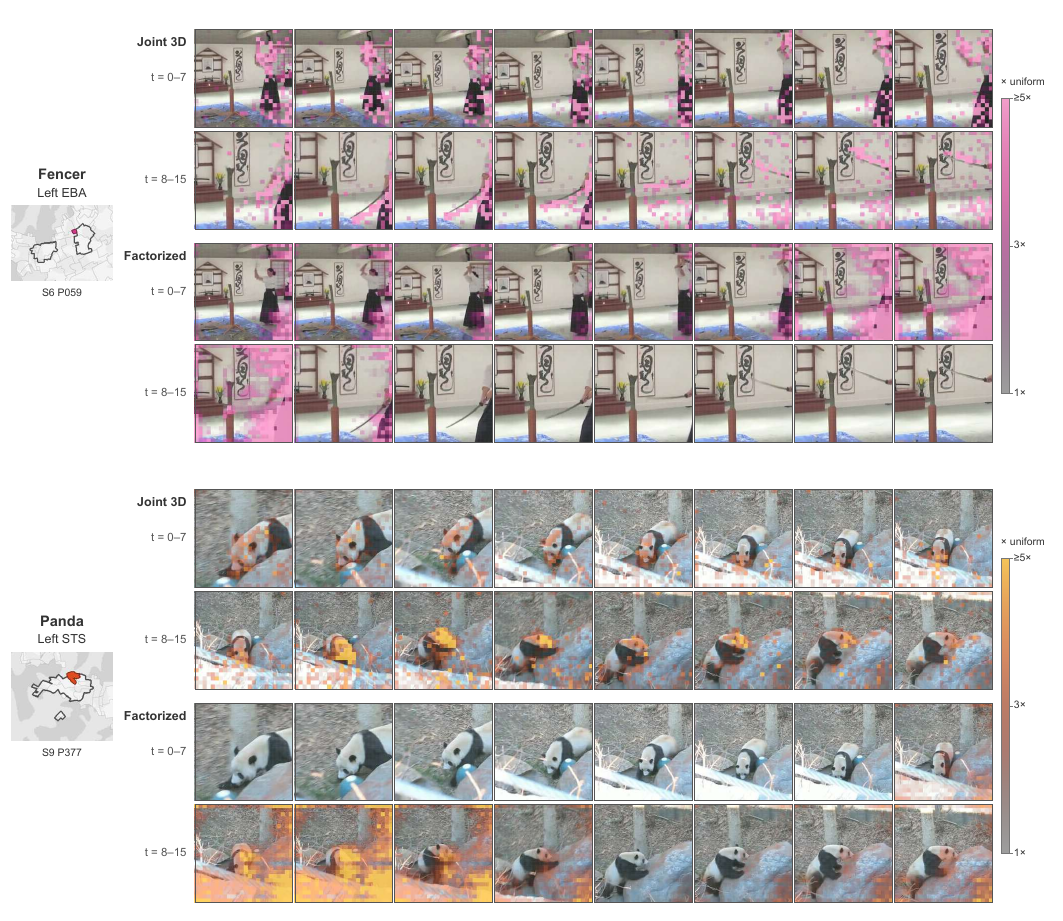}
    \vspace{-0.15in}

    \caption{Joint attention follows moving content, whereas factorized attention cannot. Attention maps of one parcel in each of two test clips, for the joint model (upper two rows of each block) and the factorized model (lower two rows), over all 16 frames of the clip. Left of each block: the subject, the parcel and its location on the flatmap within its functional ROI. Attention is shown in units of the uniform level $1/N$ on the same scale for both models: tokens at or below the uniform level are transparent, and color and opacity increase up to 5 times the uniform level.
    }
    \label{fig:fig3}
\end{figure}

\textbf{Interpreting cross-attention maps.} Fig.~\ref{fig:fig3} shows the attention maps of the joint and factorized models for one parcel in each of two test clips, both in lateral/face/body regions, where coupled routing improves accuracy most. For a parcel in left EBA, the attention of the joint model stays on the body of a fencer and, after a cut to a close-up, follows the arm and sword through the swing. For a parcel in left STS, it stays on the head and body of a panda as it climbs. The factorized model cannot: its spatial map is fixed across frames and only its temporal weights change, so it concentrates its attention on a few frames and, within them, spreads it over a large region that includes the background. In the fencer clip, a single spatial map must serve both the wide shot and the close-up, although the body is in different places in the two shots (more examples in Appendix~\ref{app:examples}).

\begin{figure}[t]
    \centering
    \includegraphics[width=0.80\linewidth]{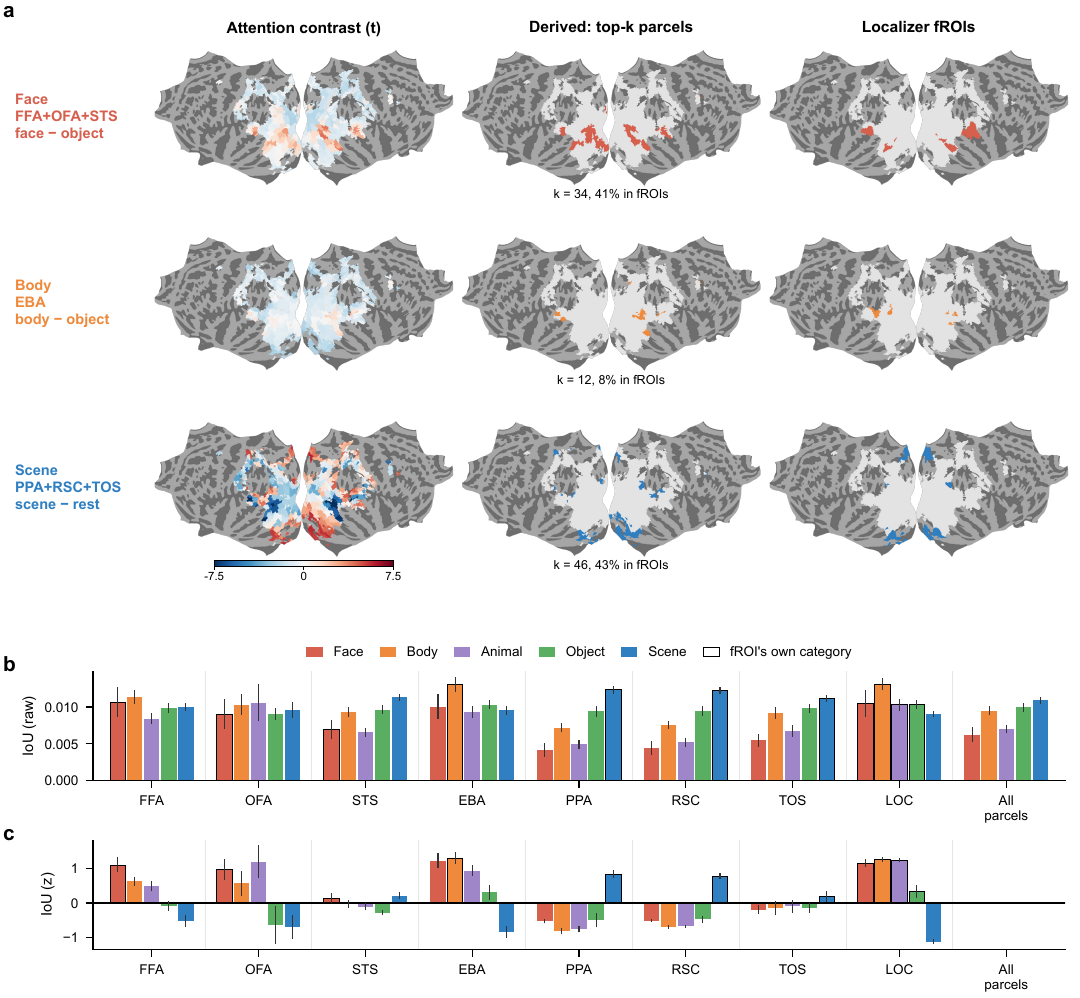}
    \vspace{-0.1in}
    \caption{The attention of the joint model differentiates category-selective regions. \textbf{(A)} Localizing category-selective regions from attention alone in one subject (S8). Left: $t$ of each localizer contrast of z-scored IoU across test clips, per parcel. Middle: the $K$ highest-$t$ parcels ($K$ = number of ROI parcels) and their overlap with the ROIs. Right: the localizer-defined ROIs (parcels with at least 20\% of their voxels inside). \textbf{(B)} Raw IoU of the 92 most-attended tokens (top 1\%) with the masks of five categories, for eight category-selective ROIs and for all parcels (mean $\pm$ SEM over 10 subjects). \textbf{(C)} The same IoU z-scored across all parcels, so that all parcels are 0 by definition. Outlined bars mark the category that defines each ROI (for LOC, all four foreground categories).
    }
    \label{fig:fig4}
\end{figure}

\textbf{Cross-attention maps differentiate category-selective networks.} If the attention of a parcel reflects what its voxels respond to, the attention of category-selective regions should land on their preferred category, and it should be possible to find these regions from the attention alone, without fMRI localizer data. We test both with the attention of the joint model on the 102 test clips, using the intersection over union (IoU) of each parcel's 92 most-attended tokens (the top 1\%) with category masks obtained by segmenting the clips with SAM~3 \citep{carion2025sam3} (Appendix~\ref{app:catsel}). Because the raw IoU mostly reflects the size of each category (Fig.~\ref{fig:fig4}b), we z-score it across parcels to better isolate parcel differences (Fig.~\ref{fig:fig4}c). This metric indicates that FFA, OFA and EBA attend more than the average parcel to faces, bodies and animals, which usually appear together in the clips, and less to scences, whereas the scene-selective PPA and RSC show the reverse pattern.
For the majority of face-, body-, and scene-selective ROIs, the highest category in this metric is the expected category  
(with exceptions OFA and STS, where animal and scene are slightly above face respectively).

Based on these z-scored IoU values, we compute ``localizer contrasts'' that quantify face, body, and scene-selectivity.  
As expected, the parcels with the highest contrast overlap with the corresponding ROIs above chance for all three localizers: face 0.19 (chance 0.09), body 0.14 (chance 0.02) and scene 0.28 (chance 0.10; two-sided one-sample $t$-tests of overlap minus chance across subjects, FDR-corrected over the three localizers, all $p_{\mathrm{FDR}} < 0.05$; Table~\ref{tab:iou-localization}). 
Additional supporting analysis can be found in Appendix~\ref{app:simple-metrics}. These results indicate that attention maps reliably differentiate the selectivity of distinct functionally-defined cortical networks.

\vspace{-0.20in}

\section{Discussion}

We developed a novel framework for modeling human fMRI responses to natural video clips, leveraging joint spatiotemporal cross-attention over embeddings from a self-supervised video model. Our results showed that joint attention outperformed spatial-only or factorized attention in multiple regions of higher visual cortex, with the most consistent advantage in lateral and parietal regions. This aligns with past evidence that dorsolateral visual cortex regions play a key role in dynamic processing of motion, events and social actions \citep{mcmahon2023, andersen1997, pitcher2011},
and expands upon these past findings by showing how cross-attention maps can yield interpretable tracking of the spatiotemporal features that drive cortical response on a per-frame basis. 

Our results could suggest that higher visual cortex implements a mechanism analogous to spatiotemporal cross-attention over feature maps during video perception, but this remains to be empirically tested. Previous work suggests that normalization across multiple objects in images \citep{bao2018representation, zoccolan2005multiple} and/or normalization across time \citep{zhou2018compressive, Groen2022} can explain neural response properties in the visual system, and it is possible that scaled dot-product attention \citep{vaswani2023attentionneed} approximates a similar normalization mechanism \citep{kozachkov2023transformers}. To evaluate this, future work should compare performance of cross-attention versus more biologically-inspired normalization models \citep{reynolds2009normalization, carandini2011normalization}.   

Other limitations of this work include the short length of the video stimuli (3s) used in the BMD dataset \citep{lahner2024bold}. Expanding our framework to modeling longer movie stimuli \citep{boyle2025cneuromod, gifford2025algonautsproject2025challenge}, would allow investigation of temporal integration over longer time windows. The slow temporal resolution of fMRI recordings presents another limitation, so future work should also explore how this model can predict data in more temporally-resolved modalities including MEG and EEG. These extensions will lead to more comprehensive understanding of how neural populations adaptively weight time-varying information during dynamic natural vision.

\newpage
\clearpage
\bibliography{iclr2027_conference}
\bibliographystyle{iclr2027_conference}

\clearpage
\section{Appendix}

\subsection{Choice of parcel-based queries}
\label{app:parcel-queries}

We assign one query to each parcel of the Schaefer 1000-parcel atlas (Section~\ref{sec:methods-data}; Appendix~\ref{app:data}). The number of queries sets a trade-off. With one query per voxel, each attention map would be estimated from a single noisy voxel; with one query per large region, a single map would be imposed on a heterogeneous population, and the regions would have to be defined in advance. Parcels sit between these extremes. Every modeled voxel belongs to exactly one parcel query, so the whole measured volume is predicted, with a single additional query for voxels without a parcel label. The parcels are fine enough for differences within and between functional regions to emerge in the attention maps. Because the atlas is defined independently of our fMRI data and functional localizers, any category selectivity that emerges in the attention maps (Appendix~\ref{app:catsel}) is not built into the model. Finally, with about 430 queries per subject, each attention map is shared by many voxels and is therefore estimated from many voxels' responses. This follows the observation that finer query granularity improves cross-attention encoding models \citep{adeli2026transformerbrainencodersexplain}, and the use of one query per Schaefer parcel in a whole-brain cross-attention encoder for static images \citep{hwang2025insilico}. Here we use the parcels in volumetric space for video, with attention over both space and time.

\subsection{Dataset and preprocessing details}
\label{app:data}

\textbf{Preprocessing and voxel selection.} We use the BMD release recommended by the dataset authors, preprocessed with fMRIPrep \citep{esteban2019fmriprep} in MNI152NLin2009cAsym volume space at 2.5\,mm. We model all voxels in the union of the dataset's group-level mask of cortex with reliable responses to the videos (``BMDGeneral'' mask) and the 23 functional ROIs. The single-trial betas were estimated by the dataset authors with GLMsingle and z-scored per voxel within each scanning session. After averaging over the repeats of each clip, we z-score each voxel again separately within the training and the test clips. The second normalization puts training and test targets on the same scale, since the test responses are averaged over more repeats (10 vs.\ 3) and are therefore less noisy. The prediction target is thus a single repeat-averaged response per voxel and clip; we do not model the response time course within a clip. The noise ceiling released with the dataset follows the procedure of the Natural Scenes Dataset \citep{allen2022massive}.

\textbf{Parcellation.} We use the volumetric MNI152 version of the Schaefer atlas, resample it to the BMD voxel grid with nearest-neighbor interpolation and intersect it with each subject's voxel mask. Each parcel with at least one voxel in the mask becomes a query, with the two hemispheres parcellated separately, and the voxels without a parcel label (mostly subcortical and edge voxels) are assigned to one additional query. Every modeled voxel therefore belongs to exactly one query.

\textbf{Functional ROIs.} The 23 ROIs are V1v, V1d, V2v, V2d, V3v, V3d, V3ab, hV4, IPS0 and IPS1-2-3 from the retinotopic atlas of \citet{wang2015probabilistic}; MT, 7AL, BA2, PFop and PFt from the HCP multimodal parcellation \citep{glasser2016multi}; and EBA, LOC, FFA, OFA, STS, PPA, RSC and TOS from the subject-specific functional localizer (bodies $>$ objects for EBA; objects $>$ scrambled objects for LOC; faces $>$ objects for FFA, OFA and STS; scenes $>$ objects for PPA, RSC and TOS).

\subsection{Implementation details}
\label{app:implementation}

\textbf{Backbone tokens.} We use the V-JEPA~2 ViT-g/16 model with 384\,px, 64-frame input. From each 3\,s clip we sample 16 frames uniformly and repeat each frame four times to fill the 64-frame input. The $2 \times 16 \times 16$ tubelets of the model give a $32 \times 24 \times 24$ token grid, and we average each pair of adjacent temporal tokens (which cover the same repeated frame), giving $16 \times 24 \times 24$ tokens of dimension $D = 1408$. Each token therefore covers one sampled frame ($\approx$0.19\,s) and a $16 \times 16$\,px patch.

\textbf{Model size.} For a typical subject the joint model has about 47M parameters, of which about 0.6M are queries, 21.8M are the shared attention and FFN block, and 24.5M are the voxel readouts.

\textbf{Data splits and ensembling.} The 102 test clips are held out and used only for the final evaluation. For each of the 10 ensemble members, the 1,000 training clips are split at random into 900 clips for training and 100 for validation, with a different split and initialization for each member, and the validation clips are used only for early stopping. Because each training clip is held out for validation in about one of the 10 members, the ensemble as a whole uses all training clips. All models, including the ridge baseline, use the same 10 splits.

\textbf{Optimization.} We use AdamW with a learning rate of $10^{-3}$, halved every 3 epochs, weight decay 0.05, batches of 8 clips, gradient-norm clipping at 1 and mixed precision. The queries are initialized from $\mathcal{N}(0, 0.02^2)$, and attention dropout of 0.1 is applied during training. Each member is trained for up to 12 epochs with early stopping on validation $R^2$ (patience 4), and the weights from the best validation epoch are restored (typically the third epoch). All hyperparameters are identical for the joint and constrained models, for all backbones and for all subjects.

\textbf{Ridge regression.} We chose mean pooling for the ridge baseline because it predicted more accurately than ridge regression on the flattened token grid after dimensionality reduction with PCA. Features are standardized using each member's training clips, and the penalty of each voxel is chosen from six values between $10^2$ and $10^7$ (log-spaced) by efficient leave-one-out cross-validation on the member's 900 training clips.

\subsection{ROI clustering for the region groups}
\label{app:roi-clustering}

To group the ROIs by function rather than by a priori anatomical assignment, we clustered the 23 ROIs by their responses to the videos. For each subject, we averaged the z-scored responses over each ROI's reliable voxels (noise ceiling $\geq$ 5\%), giving a matrix of 23 ROIs $\times$ 1,102 clips, and correlated the ROIs across clips. The correlation matrices were Fisher-z averaged over the 10 subjects and hierarchically clustered (distance $1 - r$, average linkage). Cutting the dendrogram into four clusters gives an early visual cluster (V1--V3, V3ab and hV4), a lateral/face/body cluster (OFA, FFA, EBA, MT, LOC and STS), a scene cluster (RSC, PPA, TOS, IPS0 and IPS1-2-3) and a parietal cluster (PFt, BA2, PFop and 7AL). This departs from a conventional ventral/lateral/dorsal division: FFA, OFA and LOC cluster with EBA, MT and STS rather than with the scene-selective ROIs, and IPS0 and IPS1-2-3 merge with each other first (distance $\approx$0.25) and then join the scene ROIs.

The region groups used in the main text follow these clusters except that IPS0 and IPS1-2-3 are placed in the parietal group because they lie in posterior parietal cortex; this exception is not data-driven. With the clustering assignment instead (both IPS ROIs in the scene group), only the scene and parietal groups change, and the joint model still outperforms the factorized, spatial, mean-pool and ridge models in both (two-sided paired $t$-tests across the 10 subjects, Benjamini--Hochberg FDR-corrected over these 8 tests). The differences in signed $r^2$ are $+0.0036$, $+0.0029$, $+0.0054$ and $+0.0073$ in the scene group ($t(9) = 4.08$, $2.50$, $3.69$ and $4.70$; $p_{\mathrm{FDR}} = 0.007$, $0.034$, $0.010$ and $0.005$) and $+0.0040$, $+0.0035$, $+0.0059$ and $+0.0206$ in the parietal group ($t(9) = 3.47$, $2.95$, $2.64$ and $6.87$; $p_{\mathrm{FDR}} = 0.011$, $0.022$, $0.031$ and $< 0.001$).

In the scene-selective ROIs alone (PPA, RSC and TOS), the joint model outperforms the factorized model in every subject ($+0.0041$ signed $r^2$, $t(9) = 6.16$, $p_{\mathrm{FDR}} < 0.001$, 10/10 subjects) but matches ridge regression on mean-pooled features ($+0.0007$, $t(9) = 0.42$, $p_{\mathrm{FDR}} = 0.75$, 8/10; two-sided paired $t$-tests, FDR-corrected over the 24 tests of Table~\ref{tab:fig2_ttests}). The advantage over ridge in the scene group of the clustering assignment therefore comes from the IPS ROIs, where the joint model gains substantially over ridge ($+0.018$ signed $r^2$ in both IPS0 and IPS1-2-3; $t(9) = 5.34$ and $5.55$, both $p < 0.001$, two-sided paired $t$-tests, uncorrected; 10/10 subjects). This is consistent with scene-selective cortex responding to the global layout of a scene, which is already captured by features averaged over the whole clip, so that routing to specific locations and moments adds little for these ROIs.

\subsection{Backbone and layer selection}
\label{app:layer-sweep}

We chose the backbone and layer with voxelwise ridge regression on mean-pooled features from each layer of several video and image models, fit and evaluated in the same way as the ridge baseline (Section~\ref{sec:methods-fitting}). Table~\ref{tab:layer-sweep} gives the best layer of each backbone and its whole-brain $R^2$ on the test clips, averaged over all voxels in each subject, then over the 10 subjects. The V-JEPA~2 models are the most accurate, and the three V-JEPA~2 sizes are nearly tied.

\begin{table}[h]
\centering
\footnotesize
\caption{Best layer of each backbone in the ridge-regression layer sweep (mean-pooled features; test $R^2$, mean over all voxels in 10 subjects).}
\label{tab:layer-sweep}
\begin{tabular}{lrr}
\toprule
\textbf{Backbone} & \textbf{Best layer} & \textbf{\boldmath$R^2$} \\
\midrule
V-JEPA~2 ViT-g \citep{assran2025vjepa2selfsupervisedvideo} & 32 & .134 \\
V-JEPA~2 ViT-H & 24 & .133 \\
V-JEPA~2 ViT-L & 18 & .132 \\
V-JEPA~2 ViT-L, fine-tuned on SSv2 & 18 & .125 \\
VideoMAE~v2 ViT-g \citep{wang2023videomaev2} & 20 & .109 \\
V-JEPA~2.1 ViT-g \textbf{[cite]} & 44 & .108 \\
Perception Encoder Spatial-L \citep{bolya2025perception} & 20 & .100 \\
Perception Encoder Spatial-G & 38 & .098 \\
DINOv3 ViT-7B, per frame \citep{simeoni2025dinov3} & 32 & .092 \\
\bottomrule
\end{tabular}
\end{table}

\subsection{Subject-level tests for Figure 2}
\label{app:fig2-ttests}

\begin{table}[h]
\centering
\footnotesize
\caption{Joint model minus each comparison model in region-mean signed $r^2$ on the 102 test clips (mean $\pm$ SEM over 10 subjects), with two-sided paired $t$-tests across subjects ($df = 9$). $p$ is uncorrected; $p_{\mathrm{FDR}}$ is Benjamini--Hochberg corrected over all 24 tests; values below 0.05 are in bold.  Positive $\Delta$ indicates that the joint model performed better. ``Joint better'' column indicates how many individual subjects have a numerical advantage for the joint model in each comparison. $d_z$: mean divided by the standard deviation of the per-subject differences.} 
\label{tab:fig2_ttests}
\begin{tabular}{llrrrrrr}
\toprule
\textbf{Region} & \textbf{Comparison} & \textbf{\boldmath$\Delta$ signed $r^2$} & \textbf{\boldmath$t(9)$} & \textbf{\boldmath$p$} & \textbf{\boldmath$p_{\mathrm{FDR}}$} & \textbf{Joint better} & \textbf{\boldmath$d_z$} \\
\midrule
All voxels & factorized & $+0.0043 \pm 0.0008$ & 5.39 & 4.4e-04 & \textbf{8.7e-04} & 10/10 & 1.71 \\
 & spatial & $+0.0040 \pm 0.0007$ & 5.85 & 2.4e-04 & \textbf{5.8e-04} & 10/10 & 1.85 \\
 & mean pool & $+0.0103 \pm 0.0017$ & 6.16 & 1.7e-04 & \textbf{4.5e-04} & 10/10 & 1.95 \\
 & ridge & $+0.0162 \pm 0.0019$ & 8.52 & 1.3e-05 & \textbf{1.6e-04} & 10/10 & 2.69 \\
\midrule
Early visual & factorized & $-0.0000 \pm 0.0010$ & -0.03 & 0.976 & 0.976 & 5/10 & -0.01 \\
 & spatial & $+0.0000 \pm 0.0009$ & 0.04 & 0.966 & 0.976 & 4/10 & 0.01 \\
 & mean pool & $+0.0122 \pm 0.0025$ & 4.84 & 9.2e-04 & \textbf{0.001} & 9/10 & 1.53 \\
 & ridge & $+0.0108 \pm 0.0028$ & 3.83 & 0.004 & \textbf{0.006} & 9/10 & 1.21 \\
\midrule
Lateral / face / body & factorized & $+0.0084 \pm 0.0013$ & 6.60 & 1.0e-04 & \textbf{3.4e-04} & 10/10 & 2.09 \\
 & spatial & $+0.0086 \pm 0.0011$ & 7.79 & 2.7e-05 & \textbf{1.6e-04} & 10/10 & 2.46 \\
 & mean pool & $+0.0145 \pm 0.0021$ & 6.80 & 7.9e-05 & \textbf{3.2e-04} & 10/10 & 2.15 \\
 & ridge & $+0.0216 \pm 0.0022$ & 10.04 & 3.5e-06 & \textbf{8.3e-05} & 10/10 & 3.17 \\
\midrule
Scene & factorized & $+0.0041 \pm 0.0007$ & 6.16 & 1.7e-04 & \textbf{4.5e-04} & 10/10 & 1.95 \\
 & spatial & $+0.0029 \pm 0.0011$ & 2.72 & 0.024 & \textbf{0.027} & 8/10 & 0.86 \\
 & mean pool & $+0.0033 \pm 0.0011$ & 3.07 & 0.013 & \textbf{0.017} & 8/10 & 0.97 \\
 & ridge & $+0.0007 \pm 0.0017$ & 0.42 & 0.687 & 0.749 & 8/10 & 0.13 \\
\midrule
Parietal & factorized & $+0.0037 \pm 0.0009$ & 3.99 & 0.003 & \textbf{0.005} & 9/10 & 1.26 \\
 & spatial & $+0.0034 \pm 0.0012$ & 2.93 & 0.017 & \textbf{0.020} & 9/10 & 0.93 \\
 & mean pool & $+0.0067 \pm 0.0022$ & 3.11 & 0.013 & \textbf{0.017} & 10/10 & 0.98 \\
 & ridge & $+0.0197 \pm 0.0028$ & 6.97 & 6.5e-05 & \textbf{3.1e-04} & 10/10 & 2.21 \\
\midrule
Outside ROIs & factorized & $+0.0040 \pm 0.0008$ & 4.98 & 7.6e-04 & \textbf{0.001} & 10/10 & 1.58 \\
 & spatial & $+0.0035 \pm 0.0007$ & 4.98 & 7.6e-04 & \textbf{0.001} & 10/10 & 1.58 \\
 & mean pool & $+0.0094 \pm 0.0017$ & 5.59 & 3.4e-04 & \textbf{7.4e-04} & 10/10 & 1.77 \\
 & ridge & $+0.0163 \pm 0.0020$ & 8.12 & 2.0e-05 & \textbf{1.6e-04} & 10/10 & 2.57 \\
\bottomrule
\end{tabular}
\end{table}

\subsection{Voxelwise accuracy and differences in each subject}
\label{app:flatmaps}

\begin{figure}[p]
    \centering
    \includegraphics[width=\linewidth]{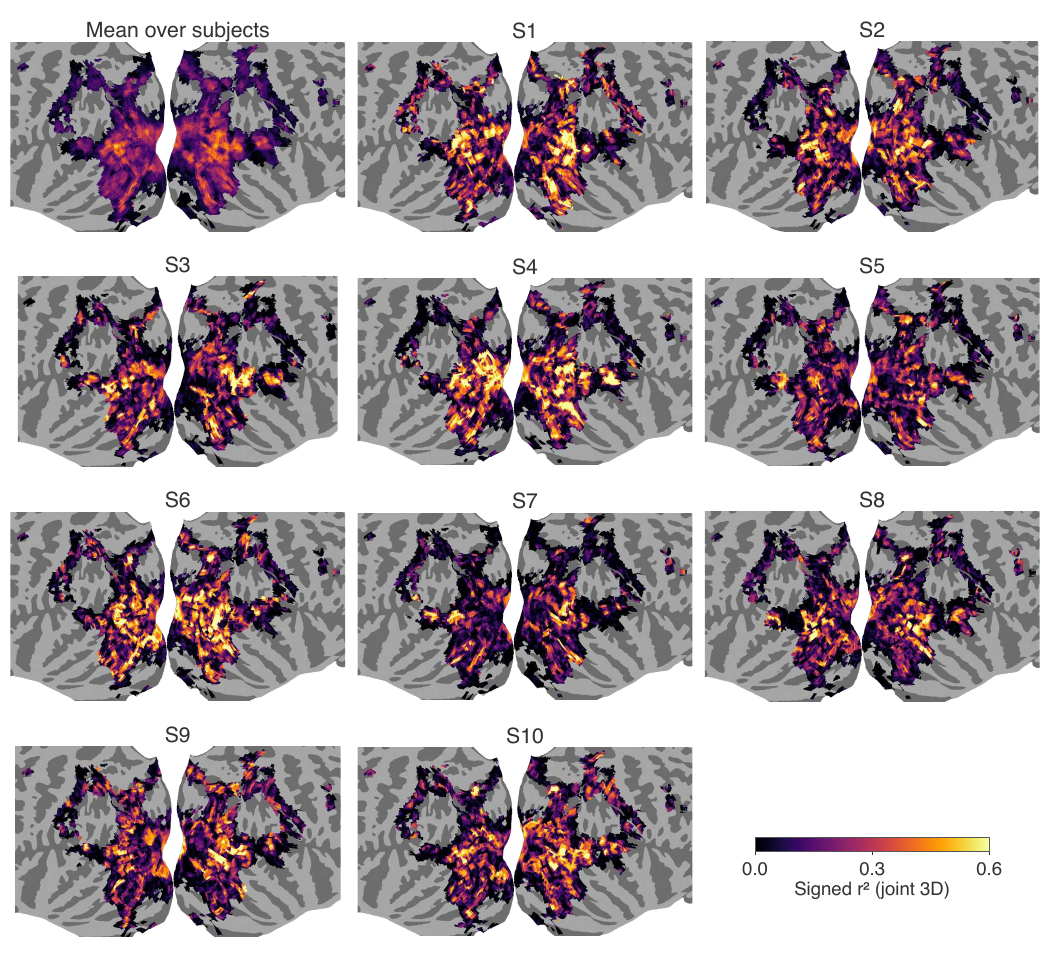}
    \caption{Voxelwise accuracy of the joint model (V-JEPA~2 ViT-g, layer 32; signed $r^2$ on the 102 test clips; all modeled voxels), averaged over the 10 subjects (top left) and in each subject, on the fsaverage flatmap. Values outside 0--0.6 are shown at the ends of the color scale.}
    \label{fig:app-flat-joint}
\end{figure}

\begin{figure}[p]
    \centering
    \includegraphics[width=\linewidth]{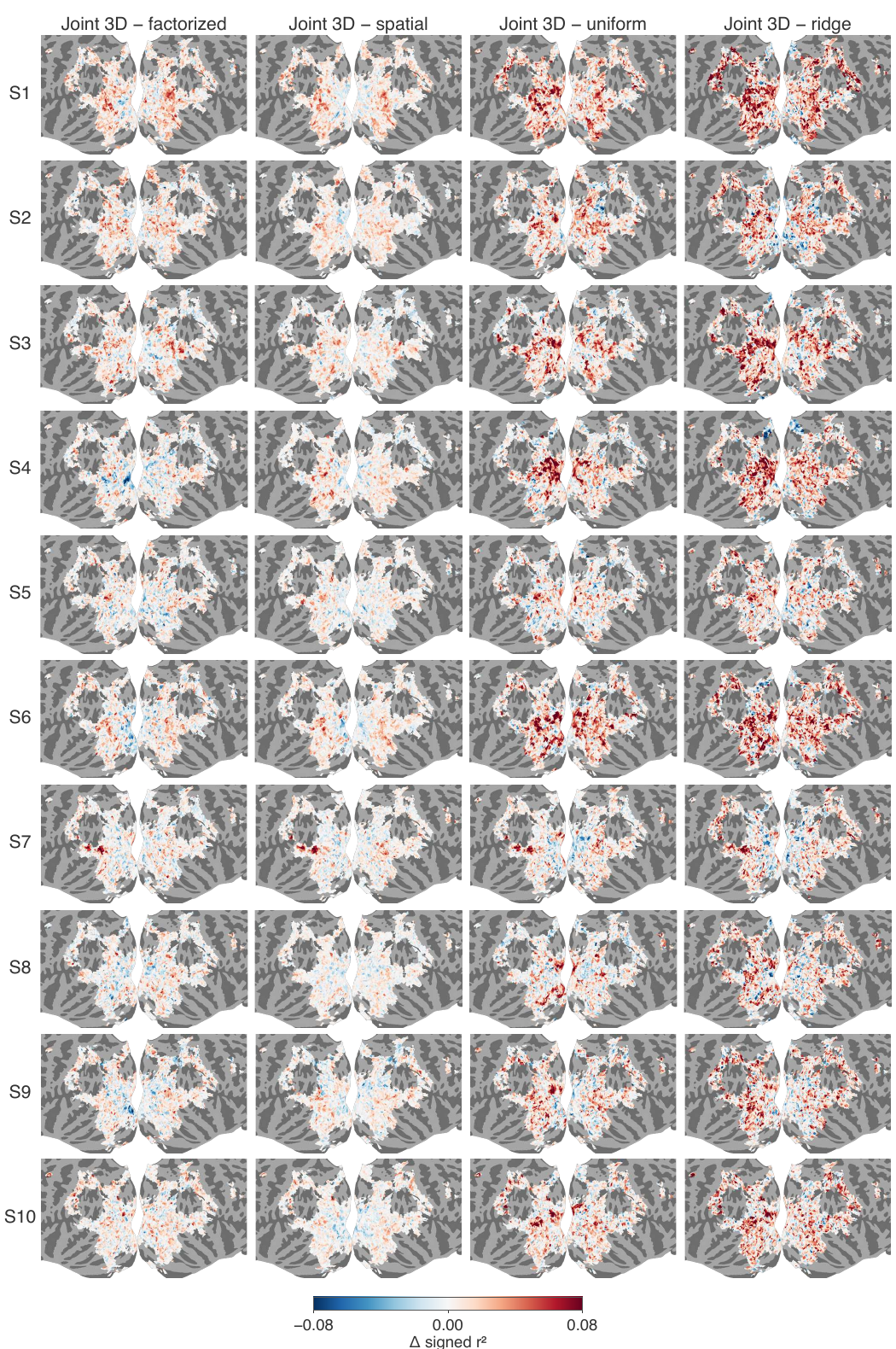}
    \caption{Joint model minus the factorized, spatial and mean-pool (``uniform'') models and ridge regression on mean-pooled features in voxelwise signed $r^2$, in each subject (rows), on the flatmap of Fig.~\ref{fig:fig2}A. All panels share one color scale; differences beyond $\pm 0.08$ are shown at its ends.}
    \label{fig:app-flat-delta}
\end{figure}

Figure~\ref{fig:app-flat-joint} shows the voxelwise accuracy of the joint model, averaged over subjects and in each subject, and Figure~\ref{fig:app-flat-delta} shows the difference between the joint model and each comparison model in each subject; Fig.~\ref{fig:fig2}A shows the average of these differences over subjects. All modeled voxels are shown, without a noise-ceiling threshold, on the fsaverage flatmap used in Fig.~\ref{fig:fig2}A. The single-subject maps are noisier than the average, but the region-level advantage of the joint model in lateral/face/body regions holds in every subject (Table~\ref{tab:fig2_ttests}).

\clearpage

\subsection{Spatial routing compared with mean pooling and ridge regression}
\label{app:spatial-routing}

Cross-attention encoders of static images improve on linear encoding models by routing each region's readout to stimulus-dependent spatial locations \citep{adeli2026transformerbrainencodersexplain}. Table~\ref{tab:spatial-tests} tests the video analogue of this result, the spatial model (temporal factor uniform, spatial factor routed), against the mean-pool model and against ridge regression on mean-pooled features, for every backbone and layer of the routing sweep (Appendix~\ref{app:routing-sweep}). The spatial model is more accurate than both in all regions for every configuration except in scene-selective cortex, where it is not significantly different from either for V-JEPA~2 layer 32 and VideoMAE~v2 layer 20, and for the last stage of CLIP, where the spatial model is less accurate than ridge regression in four of the six regions and less accurate than mean pooling in lateral/face/body regions. Of the 96 tests, 83 favor the spatial model at $p_{\mathrm{FDR}} < 0.05$.

\begin{table}[h]
\centering
\scriptsize
\setlength{\tabcolsep}{4pt}
\caption{Spatial model minus the mean-pool model and minus ridge regression on mean-pooled features, in region-mean signed $r^2$ on the 102 test clips ($\Delta$ in units of $10^{-3}$, mean over 10 subjects; voxels with noise ceiling $\geq$ 5\%). $t(9)$: two-sided paired $t$-test across subjects. $p_{\mathrm{FDR}}$: Benjamini--Hochberg corrected over all 96 tests in the table; values below 0.05 are in bold. $n_+$: subjects in which the spatial model is better.}
\label{tab:spatial-tests}
\begin{tabular}{lrrrrrrrr}
\toprule
 & \multicolumn{4}{c}{\textbf{Spatial $-$ mean pool}} & \multicolumn{4}{c}{\textbf{Spatial $-$ ridge}} \\
\cmidrule(lr){2-5} \cmidrule(lr){6-9}
\textbf{Region} & \textbf{\boldmath$\Delta$} & \textbf{\boldmath$t(9)$} & \textbf{\boldmath$p_{\mathrm{FDR}}$} & \textbf{\boldmath$n_+$} & \textbf{\boldmath$\Delta$} & \textbf{\boldmath$t(9)$} & \textbf{\boldmath$p_{\mathrm{FDR}}$} & \textbf{\boldmath$n_+$} \\
\midrule
\multicolumn{9}{l}{\textit{V-JEPA~2 ViT-g, layer 32 of 40}} \\
All voxels & $+6.3$ & 5.07 & \textbf{8.9e-4} & 9/10 & $+12.2$ & 8.26 & \textbf{3.3e-5} & 10/10 \\
Early visual & $+12.2$ & 5.58 & \textbf{4.7e-4} & 10/10 & $+10.8$ & 4.24 & \textbf{0.003} & 9/10 \\
Lateral/face/body & $+5.9$ & 3.97 & \textbf{0.004} & 9/10 & $+13.0$ & 8.73 & \textbf{2.5e-5} & 10/10 \\
Scene & $+0.5$ & 0.37 & 0.734 & 6/10 & $-2.2$ & -1.11 & 0.313 & 5/10 \\
Parietal & $+3.3$ & 2.38 & \textbf{0.045} & 7/10 & $+16.3$ & 8.09 & \textbf{3.7e-5} & 10/10 \\
Outside ROIs & $+5.9$ & 4.60 & \textbf{0.002} & 9/10 & $+12.8$ & 8.25 & \textbf{3.3e-5} & 10/10 \\
\midrule
\multicolumn{9}{l}{\textit{V-JEPA~2 ViT-g, layer 4 of 40}} \\
All voxels & $+14.3$ & 11.29 & \textbf{5.7e-6} & 10/10 & $+18.4$ & 18.77 & \textbf{1.3e-6} & 10/10 \\
Early visual & $+18.7$ & 9.78 & \textbf{1.2e-5} & 10/10 & $+18.2$ & 11.80 & \textbf{4.9e-6} & 10/10 \\
Lateral/face/body & $+14.7$ & 9.80 & \textbf{1.2e-5} & 10/10 & $+21.2$ & 14.05 & \textbf{2.3e-6} & 10/10 \\
Scene & $+10.4$ & 5.88 & \textbf{3.4e-4} & 10/10 & $+14.1$ & 8.34 & \textbf{3.2e-5} & 10/10 \\
Parietal & $+7.9$ & 8.21 & \textbf{3.4e-5} & 10/10 & $+12.2$ & 11.81 & \textbf{4.9e-6} & 10/10 \\
Outside ROIs & $+14.4$ & 9.94 & \textbf{1.2e-5} & 10/10 & $+18.9$ & 16.60 & \textbf{1.4e-6} & 10/10 \\
\midrule
\multicolumn{9}{l}{\textit{VideoMAE~v2 ViT-g, layer 20 of 40}} \\
All voxels & $+14.8$ & 8.91 & \textbf{2.3e-5} & 10/10 & $+24.6$ & 10.90 & \textbf{6.9e-6} & 10/10 \\
Early visual & $+20.6$ & 6.36 & \textbf{1.9e-4} & 10/10 & $+23.4$ & 5.22 & \textbf{7.3e-4} & 10/10 \\
Lateral/face/body & $+19.6$ & 12.41 & \textbf{4.0e-6} & 10/10 & $+34.5$ & 16.20 & \textbf{1.4e-6} & 10/10 \\
Scene & $+2.0$ & 1.61 & 0.152 & 8/10 & $+1.4$ & 0.76 & 0.483 & 5/10 \\
Parietal & $+8.5$ & 4.81 & \textbf{0.001} & 10/10 & $+21.6$ & 7.37 & \textbf{7.1e-5} & 10/10 \\
Outside ROIs & $+13.6$ & 7.27 & \textbf{7.8e-5} & 10/10 & $+23.7$ & 10.00 & \textbf{1.1e-5} & 10/10 \\
\midrule
\multicolumn{9}{l}{\textit{VideoMAE~v2 ViT-g, layer 4 of 40}} \\
All voxels & $+19.3$ & 8.25 & \textbf{3.3e-5} & 10/10 & $+25.6$ & 9.45 & \textbf{1.5e-5} & 10/10 \\
Early visual & $+32.4$ & 6.40 & \textbf{1.9e-4} & 10/10 & $+30.6$ & 5.76 & \textbf{3.8e-4} & 10/10 \\
Lateral/face/body & $+22.4$ & 8.70 & \textbf{2.5e-5} & 10/10 & $+33.2$ & 12.43 & \textbf{4.0e-6} & 10/10 \\
Scene & $+10.2$ & 6.14 & \textbf{2.5e-4} & 10/10 & $+17.6$ & 9.01 & \textbf{2.1e-5} & 10/10 \\
Parietal & $+8.6$ & 8.54 & \textbf{2.8e-5} & 10/10 & $+16.1$ & 14.19 & \textbf{2.3e-6} & 10/10 \\
Outside ROIs & $+17.3$ & 7.21 & \textbf{8.2e-5} & 10/10 & $+23.9$ & 8.68 & \textbf{2.5e-5} & 10/10 \\
\midrule
\multicolumn{9}{l}{\textit{DINOv2 ViT-B/14, block 12 of 12}} \\
All voxels & $+22.9$ & 10.71 & \textbf{7.2e-6} & 10/10 & $+27.1$ & 9.85 & \textbf{1.2e-5} & 10/10 \\
Early visual & $+32.0$ & 8.52 & \textbf{2.8e-5} & 10/10 & $+31.0$ & 6.79 & \textbf{1.3e-4} & 10/10 \\
Lateral/face/body & $+27.2$ & 13.79 & \textbf{2.3e-6} & 10/10 & $+34.2$ & 14.33 & \textbf{2.3e-6} & 10/10 \\
Scene & $+9.4$ & 11.54 & \textbf{5.1e-6} & 10/10 & $+8.6$ & 4.50 & \textbf{0.002} & 10/10 \\
Parietal & $+11.5$ & 5.46 & \textbf{5.4e-4} & 10/10 & $+19.0$ & 7.02 & \textbf{9.9e-5} & 10/10 \\
Outside ROIs & $+22.0$ & 9.59 & \textbf{1.4e-5} & 10/10 & $+26.4$ & 8.81 & \textbf{2.4e-5} & 10/10 \\
\midrule
\multicolumn{9}{l}{\textit{DINOv2 ViT-B/14, block 2 of 12}} \\
All voxels & $+23.0$ & 10.77 & \textbf{7.1e-6} & 10/10 & $+25.5$ & 13.75 & \textbf{2.3e-6} & 10/10 \\
Early visual & $+51.6$ & 11.75 & \textbf{4.9e-6} & 10/10 & $+40.1$ & 10.58 & \textbf{7.7e-6} & 10/10 \\
Lateral/face/body & $+20.4$ & 10.80 & \textbf{7.1e-6} & 10/10 & $+29.1$ & 17.66 & \textbf{1.3e-6} & 10/10 \\
Scene & $+6.7$ & 4.26 & \textbf{0.003} & 9/10 & $+14.5$ & 6.48 & \textbf{1.7e-4} & 10/10 \\
Parietal & $+5.5$ & 6.65 & \textbf{1.5e-4} & 10/10 & $+11.5$ & 12.45 & \textbf{4.0e-6} & 10/10 \\
Outside ROIs & $+20.8$ & 8.83 & \textbf{2.4e-5} & 10/10 & $+23.6$ & 12.01 & \textbf{4.9e-6} & 10/10 \\
\midrule
\multicolumn{9}{l}{\textit{CLIP RN50, stage 4 of 4}} \\
All voxels & $+0.0$ & 0.01 & 0.991 & 5/10 & $-6.7$ & -2.73 & \textbf{0.026} & 3/10 \\
Early visual & $+15.7$ & 4.15 & \textbf{0.003} & 9/10 & $+10.0$ & 2.65 & \textbf{0.029} & 8/10 \\
Lateral/face/body & $-15.2$ & -3.46 & \textbf{0.008} & 1/10 & $-22.2$ & -5.78 & \textbf{3.8e-4} & 0/10 \\
Scene & $+6.5$ & 4.20 & \textbf{0.003} & 10/10 & $-5.3$ & -4.14 & \textbf{0.003} & 2/10 \\
Parietal & $+2.6$ & 0.96 & 0.378 & 7/10 & $-2.9$ & -1.30 & 0.241 & 4/10 \\
Outside ROIs & $+0.3$ & 0.13 & 0.910 & 5/10 & $-6.4$ & -2.97 & \textbf{0.018} & 3/10 \\
\midrule
\multicolumn{9}{l}{\textit{CLIP RN50, stage 2 of 4}} \\
All voxels & $+28.7$ & 9.29 & \textbf{1.7e-5} & 10/10 & $+41.9$ & 11.62 & \textbf{5.1e-6} & 10/10 \\
Early visual & $+50.7$ & 7.48 & \textbf{6.5e-5} & 10/10 & $+54.8$ & 7.56 & \textbf{6.0e-5} & 10/10 \\
Lateral/face/body & $+35.0$ & 12.71 & \textbf{4.0e-6} & 10/10 & $+53.8$ & 15.14 & \textbf{2.0e-6} & 10/10 \\
Scene & $+5.2$ & 3.35 & \textbf{0.010} & 10/10 & $+17.8$ & 11.27 & \textbf{5.7e-6} & 10/10 \\
Parietal & $+12.3$ & 7.65 & \textbf{5.6e-5} & 10/10 & $+24.6$ & 10.38 & \textbf{8.7e-6} & 10/10 \\
Outside ROIs & $+25.4$ & 8.04 & \textbf{3.9e-5} & 10/10 & $+39.1$ & 10.95 & \textbf{6.9e-6} & 10/10 \\
\bottomrule
\end{tabular}
\end{table}

\subsection{Contributions of the readout components}
\label{app:ablation}

\begin{figure}[p]
    \centering
    \includegraphics[width=\linewidth]{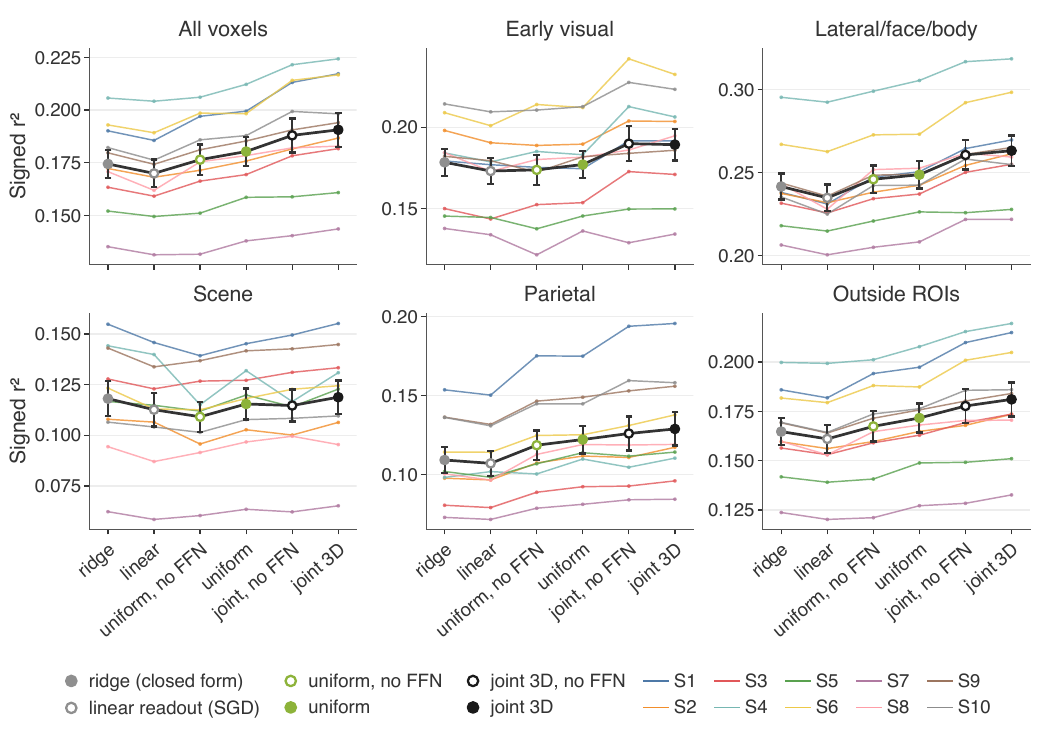}
    \caption{Contributions of the readout components (V-JEPA~2 ViT-g, layer 32; region-mean signed $r^2$ on the 102 test clips; voxels with noise ceiling $\geq$ 5\%). From left to right: ridge regression on mean-pooled features, a linear readout of the mean-pooled features trained by gradient descent, the mean-pool model without and with the FFN (``uniform''), and the joint model without and with the FFN. Open markers: models without an FFN. Colored lines: individual subjects; markers: mean $\pm$ SEM over subjects.}
    \label{fig:app-ablation}
\end{figure}

Besides routing, the joint model differs from ridge regression on mean-pooled features in three ways: it is trained by gradient descent, it passes the tokens through value and output projections shared by all voxels, and it applies a feedforward network (FFN) to each parcel's representation before the voxel readouts. To separate these components, we trained three further models on V-JEPA~2 ViT-g layer 32, with the same data splits, ensembling, optimization and early stopping as the joint model (Appendix~\ref{app:implementation}):
\begin{itemize}
    \item \textbf{Linear readout:} the voxel readouts applied directly to the mean-pooled tokens and trained by gradient descent, the counterpart of ridge regression.
    \item \textbf{Uniform, no FFN:} the mean-pool model without the FFN, so that every parcel's representation is the output projection of the mean of the value projections. Because both projections are linear, this model is still linear in the mean-pooled tokens; it differs from the linear readout only in its parametrization.
    \item \textbf{Joint, no FFN:} the joint model without the FFN.
\end{itemize}
With ridge regression, the mean-pool model and the joint model, these form a sequence that adds one component at a time (Fig.~\ref{fig:app-ablation}, Table~\ref{tab:ablation-tests}; two-sided paired $t$-tests across the 10 subjects, $df = 9$, FDR-corrected over the 30 tests of the table). Routing contributes the most. Without the FFN, joint routing improves on uniform attention in all six regions and in every subject ($+0.0115$ across all voxels; $t(9) = 10.51$, $p_{\mathrm{FDR}} < 0.001$), as much as it does with the FFN (joint minus mean pool, $+0.0103$; Table~\ref{tab:fig2_ttests}). The FFN adds less, with uniform attention ($+0.0039$; $t(9) = 5.29$, $p_{\mathrm{FDR}} = 0.001$) and with joint routing ($+0.0026$; $t(9) = 4.74$, $p_{\mathrm{FDR}} = 0.002$), and in early visual cortex it adds nothing significant in either case. Gradient descent by itself does not explain the advantage of the attention models: the linear readout is less accurate than ridge regression in every region ($-0.0045$ across all voxels; $t(9) = -7.11$, $p_{\mathrm{FDR}} < 0.001$). The shared projections make up this deficit and more in lateral/face/body and parietal regions, but not in early visual or scene regions.

\begin{table}[!htbp]
\centering
\scriptsize
\setlength{\tabcolsep}{4pt}
\caption{Each model of Fig.~\ref{fig:app-ablation} minus the model it extends, in region-mean signed $r^2$ (mean $\pm$ SEM over 10 subjects), with two-sided paired $t$-tests across subjects ($df = 9$). $p$ is uncorrected; $p_{\mathrm{FDR}}$ is Benjamini--Hochberg corrected over all 30 tests; values below 0.05 are in bold. $n_+$: subjects in which the first model is better.}
\label{tab:ablation-tests}
\begin{tabular}{llrrrrr}
\toprule
\textbf{Contrast} & \textbf{Region} & \textbf{\boldmath$\Delta$ signed $r^2$} & \textbf{\boldmath$t(9)$} & \textbf{\boldmath$p$} & \textbf{\boldmath$p_{\mathrm{FDR}}$} & \textbf{\boldmath$n_+$} \\
\midrule
Linear readout (SGD) $-$ ridge & All voxels & $-0.0045 \pm 0.0006$ & -7.11 & 5.6e-05 & \textbf{4.2e-04} & 0/10 \\
 & Early visual & $-0.0053 \pm 0.0010$ & -5.37 & 4.5e-04 & \textbf{0.001} & 0/10 \\
 & Lateral/face/body & $-0.0069 \pm 0.0012$ & -5.88 & 2.3e-04 & \textbf{0.001} & 0/10 \\
 & Scene & $-0.0055 \pm 0.0010$ & -5.26 & 5.2e-04 & \textbf{0.001} & 0/10 \\
 & Parietal & $-0.0021 \pm 0.0008$ & -2.53 & 0.032 & \textbf{0.039} & 1/10 \\
 & Outside ROIs & $-0.0037 \pm 0.0006$ & -6.71 & 8.8e-05 & \textbf{4.4e-04} & 0/10 \\
\midrule
Uniform, no FFN $-$ linear readout & All voxels & $+0.0065 \pm 0.0014$ & 4.53 & 0.001 & \textbf{0.002} & 10/10 \\
 & Early visual & $+0.0007 \pm 0.0026$ & 0.27 & 0.790 & 0.790 & 5/10 \\
 & Lateral/face/body & $+0.0112 \pm 0.0020$ & 5.72 & 2.9e-04 & \textbf{0.001} & 10/10 \\
 & Scene & $-0.0035 \pm 0.0028$ & -1.24 & 0.248 & 0.265 & 4/10 \\
 & Parietal & $+0.0115 \pm 0.0022$ & 5.33 & 4.8e-04 & \textbf{0.001} & 9/10 \\
 & Outside ROIs & $+0.0063 \pm 0.0013$ & 4.76 & 0.001 & \textbf{0.002} & 10/10 \\
\midrule
Uniform $-$ uniform, no FFN & All voxels & $+0.0039 \pm 0.0007$ & 5.29 & 5.0e-04 & \textbf{0.001} & 9/10 \\
 & Early visual & $+0.0033 \pm 0.0017$ & 1.88 & 0.093 & 0.104 & 7/10 \\
 & Lateral/face/body & $+0.0028 \pm 0.0007$ & 3.93 & 0.003 & \textbf{0.005} & 10/10 \\
 & Scene & $+0.0063 \pm 0.0013$ & 4.70 & 0.001 & \textbf{0.002} & 10/10 \\
 & Parietal & $+0.0036 \pm 0.0010$ & 3.51 & 0.007 & \textbf{0.009} & 8/10 \\
 & Outside ROIs & $+0.0043 \pm 0.0008$ & 5.50 & 3.8e-04 & \textbf{0.001} & 9/10 \\
\midrule
Joint, no FFN $-$ uniform, no FFN & All voxels & $+0.0115 \pm 0.0011$ & 10.51 & 2.4e-06 & \textbf{3.3e-05} & 10/10 \\
 & Early visual & $+0.0161 \pm 0.0024$ & 6.77 & 8.2e-05 & \textbf{4.4e-04} & 10/10 \\
 & Lateral/face/body & $+0.0147 \pm 0.0014$ & 10.66 & 2.1e-06 & \textbf{3.3e-05} & 10/10 \\
 & Scene & $+0.0055 \pm 0.0010$ & 5.38 & 4.4e-04 & \textbf{0.001} & 10/10 \\
 & Parietal & $+0.0074 \pm 0.0016$ & 4.59 & 0.001 & \textbf{0.002} & 10/10 \\
 & Outside ROIs & $+0.0103 \pm 0.0010$ & 10.09 & 3.3e-06 & \textbf{3.3e-05} & 10/10 \\
\midrule
Joint $-$ joint, no FFN & All voxels & $+0.0026 \pm 0.0006$ & 4.74 & 0.001 & \textbf{0.002} & 9/10 \\
 & Early visual & $-0.0006 \pm 0.0017$ & -0.37 & 0.720 & 0.744 & 4/10 \\
 & Lateral/face/body & $+0.0026 \pm 0.0011$ & 2.29 & 0.048 & 0.055 & 8/10 \\
 & Scene & $+0.0041 \pm 0.0016$ & 2.62 & 0.028 & \textbf{0.035} & 9/10 \\
 & Parietal & $+0.0029 \pm 0.0009$ & 3.34 & 0.009 & \textbf{0.011} & 9/10 \\
 & Outside ROIs & $+0.0033 \pm 0.0006$ & 5.55 & 3.6e-04 & \textbf{0.001} & 10/10 \\
\bottomrule
\end{tabular}
\end{table}

\clearpage

\subsection{Routing in each functional ROI}
\label{app:roi-routing}

\begin{figure}[p]
    \centering
    \includegraphics[width=\linewidth]{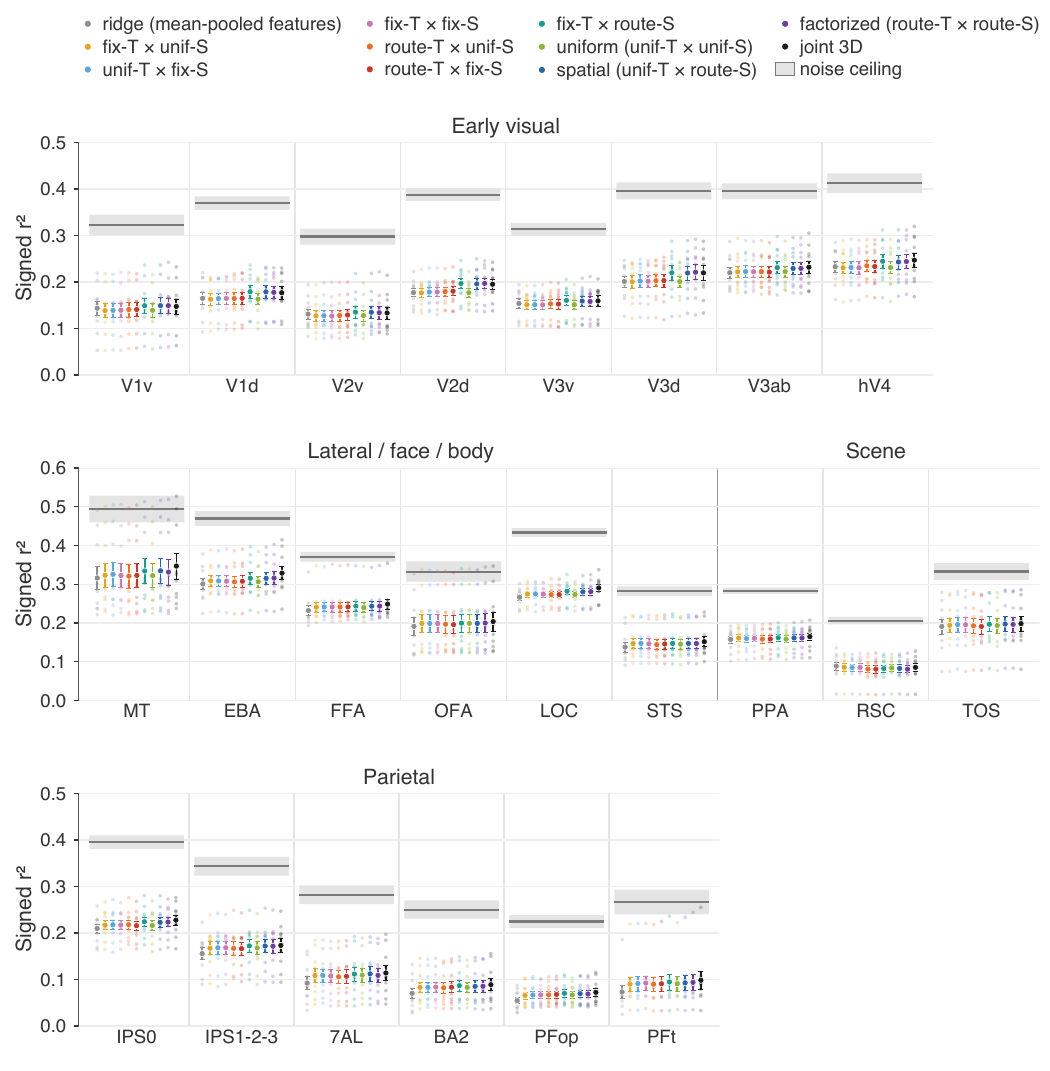}
    \caption{Region-mean signed $r^2$ of every routing configuration, the joint model and ridge regression in each functional ROI (V-JEPA~2 ViT-g, layer 32; 102 test clips; voxels with noise ceiling $\geq$ 5\%), grouped by region as in Fig.~\ref{fig:fig2}. Large dots: mean $\pm$ SEM over subjects; small dots: individual subjects; grey band: noise ceiling (mean $\pm$ SEM).}
    \label{fig:app-roi-routing}
\end{figure}

Figure~\ref{fig:app-roi-routing} repeats Fig.~\ref{fig:fig2}B for each of the 23 functional ROIs, and Table~\ref{tab:roi-tests} compares the joint model with the factorized, spatial and mean-pool models and ridge regression in each ROI (two-sided paired $t$-tests across the 10 subjects, $df = 9$, Benjamini--Hochberg FDR-corrected over the 92 tests of the table). The ROIs follow the pattern of their region groups. In each of the eight early visual ROIs, the joint model does not differ significantly from the factorized or the spatial model, and it is better than mean pooling in seven of them. In the lateral/face/body group, it is better than the spatial model in all six ROIs and than the factorized model in all but OFA ($p_{\mathrm{FDR}} = 0.054$), with the largest gains over the factorized model in MT ($+0.016$; $t(9) = 6.54$, $p_{\mathrm{FDR}} < 0.001$) and EBA ($+0.013$; $t(9) = 8.50$, $p_{\mathrm{FDR}} < 0.001$). Among the scene-selective ROIs, the joint model is better than every comparison model in PPA, but not significantly better than any of them in TOS, and in RSC it is better only than the factorized model.

\begin{table}[!htbp]
\centering
\scriptsize
\setlength{\tabcolsep}{2.6pt}
\caption{Joint model minus each comparison model in ROI-mean signed $r^2$ on the 102 test clips ($\Delta$ in units of $10^{-3}$, mean over 10 subjects; voxels with noise ceiling $\geq$ 5\%). $t(9)$: two-sided paired $t$-test across subjects. $p_{\mathrm{FDR}}$: Benjamini--Hochberg corrected over all 92 tests in the table; values below 0.05 are in bold. $n_+$: subjects (of 10) in which the joint model is better.}
\label{tab:roi-tests}
\begin{tabular}{lrrrrrrrrrrrrrrrr}
\toprule
 & \multicolumn{4}{c}{\textbf{Joint $-$ factorized}} & \multicolumn{4}{c}{\textbf{Joint $-$ spatial}} & \multicolumn{4}{c}{\textbf{Joint $-$ mean pool}} & \multicolumn{4}{c}{\textbf{Joint $-$ ridge}} \\
\cmidrule(lr){2-5} \cmidrule(lr){6-9} \cmidrule(lr){10-13} \cmidrule(lr){14-17}
\textbf{ROI} & \textbf{\boldmath$\Delta$} & \textbf{\boldmath$t(9)$} & \textbf{\boldmath$p_{\mathrm{FDR}}$} & \textbf{\boldmath$n_+$} & \textbf{\boldmath$\Delta$} & \textbf{\boldmath$t(9)$} & \textbf{\boldmath$p_{\mathrm{FDR}}$} & \textbf{\boldmath$n_+$} & \textbf{\boldmath$\Delta$} & \textbf{\boldmath$t(9)$} & \textbf{\boldmath$p_{\mathrm{FDR}}$} & \textbf{\boldmath$n_+$} & \textbf{\boldmath$\Delta$} & \textbf{\boldmath$t(9)$} & \textbf{\boldmath$p_{\mathrm{FDR}}$} & \textbf{\boldmath$n_+$} \\
\midrule
\multicolumn{17}{l}{\textit{Early visual}} \\
V1v & $-1.7$ & -1.26 & 0.281 & 5 & $-1.7$ & -1.53 & 0.203 & 3 & $+8.3$ & 3.38 & \textbf{0.017} & 8 & $+3.8$ & 1.40 & 0.240 & 6 \\
V1d & $-0.8$ & -0.59 & 0.590 & 4 & $-1.9$ & -2.26 & 0.072 & 3 & $+13.9$ & 4.30 & \textbf{0.006} & 9 & $+12.0$ & 3.63 & \textbf{0.013} & 8 \\
V2v & $-0.5$ & -0.40 & 0.717 & 4 & $-1.5$ & -1.84 & 0.130 & 4 & $+5.5$ & 2.18 & 0.080 & 6 & $+2.5$ & 0.95 & 0.402 & 7 \\
V2d & $-1.8$ & -1.23 & 0.290 & 4 & $-0.7$ & -0.68 & 0.546 & 4 & $+18.3$ & 5.92 & \textbf{0.001} & 9 & $+18.0$ & 5.38 & \textbf{0.002} & 9 \\
V3v & $+1.2$ & 0.92 & 0.412 & 6 & $+0.3$ & 0.31 & 0.772 & 5 & $+8.3$ & 4.59 & \textbf{0.004} & 9 & $+5.6$ & 3.01 & \textbf{0.026} & 8 \\
V3d & $-1.6$ & -1.10 & 0.339 & 4 & $+0.2$ & 0.26 & 0.803 & 6 & $+18.7$ & 4.08 & \textbf{0.007} & 8 & $+18.5$ & 3.51 & \textbf{0.015} & 9 \\
V3ab & $+2.5$ & 1.44 & 0.229 & 7 & $+2.7$ & 1.62 & 0.177 & 8 & $+9.5$ & 3.22 & \textbf{0.021} & 8 & $+11.7$ & 3.34 & \textbf{0.018} & 8 \\
hV4 & $+3.0$ & 1.95 & 0.112 & 8 & $+3.9$ & 2.25 & 0.072 & 8 & $+16.0$ & 5.14 & \textbf{0.002} & 10 & $+14.0$ & 3.47 & \textbf{0.016} & 10 \\
\midrule
\multicolumn{17}{l}{\textit{Lateral / face / body}} \\
MT & $+15.6$ & 6.54 & \textbf{9.8e-4} & 10 & $+12.3$ & 6.22 & \textbf{0.001} & 10 & $+24.7$ & 6.27 & \textbf{0.001} & 10 & $+30.7$ & 8.31 & \textbf{2.8e-4} & 10 \\
EBA & $+12.7$ & 8.50 & \textbf{2.8e-4} & 10 & $+13.3$ & 8.19 & \textbf{2.8e-4} & 10 & $+22.3$ & 5.73 & \textbf{0.001} & 10 & $+27.9$ & 7.71 & \textbf{3.4e-4} & 10 \\
FFA & $+5.1$ & 3.29 & \textbf{0.019} & 8 & $+5.3$ & 3.70 & \textbf{0.012} & 9 & $+9.1$ & 4.32 & \textbf{0.006} & 9 & $+16.6$ & 8.16 & \textbf{2.8e-4} & 10 \\
OFA & $+3.7$ & 2.49 & 0.054 & 8 & $+4.7$ & 4.06 & \textbf{0.007} & 10 & $+4.8$ & 3.28 & \textbf{0.019} & 8 & $+12.7$ & 4.15 & \textbf{0.007} & 9 \\
LOC & $+9.4$ & 4.86 & \textbf{0.003} & 9 & $+9.9$ & 6.30 & \textbf{0.001} & 10 & $+16.9$ & 6.17 & \textbf{0.001} & 10 & $+24.0$ & 8.87 & \textbf{2.8e-4} & 10 \\
STS & $+4.7$ & 5.00 & \textbf{0.003} & 10 & $+4.4$ & 5.68 & \textbf{0.001} & 10 & $+6.2$ & 5.81 & \textbf{0.001} & 10 & $+14.0$ & 11.67 & \textbf{9.0e-5} & 10 \\
\midrule
\multicolumn{17}{l}{\textit{Scene}} \\
PPA & $+3.5$ & 4.54 & \textbf{0.005} & 10 & $+3.5$ & 3.02 & \textbf{0.026} & 7 & $+6.7$ & 4.41 & \textbf{0.005} & 9 & $+7.3$ & 3.44 & \textbf{0.016} & 9 \\
RSC & $+4.5$ & 3.93 & \textbf{0.009} & 9 & $+2.7$ & 2.54 & 0.051 & 8 & $+1.3$ & 0.96 & 0.402 & 5 & $-3.4$ & -1.72 & 0.156 & 4 \\
TOS & $+2.0$ & 1.04 & 0.365 & 6 & $+1.1$ & 0.60 & 0.587 & 7 & $+4.7$ & 1.19 & 0.303 & 8 & $+7.4$ & 2.41 & 0.060 & 8 \\
\midrule
\multicolumn{17}{l}{\textit{Parietal}} \\
IPS0 & $+4.1$ & 1.93 & 0.114 & 7 & $+4.6$ & 2.98 & \textbf{0.027} & 9 & $+11.4$ & 4.23 & \textbf{0.006} & 10 & $+17.9$ & 5.34 & \textbf{0.002} & 10 \\
IPS1-2-3 & $+1.7$ & 1.32 & 0.264 & 5 & $+1.2$ & 0.74 & 0.510 & 6 & $+5.3$ & 1.98 & 0.108 & 7 & $+17.5$ & 5.55 & \textbf{0.002} & 10 \\
7AL & $+5.3$ & 3.72 & \textbf{0.012} & 9 & $+1.9$ & 1.33 & 0.261 & 5 & $+4.6$ & 2.53 & 0.051 & 8 & $+21.8$ & 8.04 & \textbf{2.8e-4} & 10 \\
BA2 & $+3.3$ & 2.35 & 0.066 & 8 & $+3.6$ & 3.14 & \textbf{0.022} & 9 & $+5.7$ & 2.25 & 0.072 & 7 & $+18.6$ & 6.27 & \textbf{0.001} & 10 \\
PFop & $+3.5$ & 2.67 & \textbf{0.042} & 9 & $+2.9$ & 2.97 & \textbf{0.027} & 9 & $+5.6$ & 3.17 & \textbf{0.022} & 8 & $+17.8$ & 7.49 & \textbf{3.8e-4} & 10 \\
PFt & $+4.5$ & 3.11 & \textbf{0.023} & 8 & $+5.5$ & 2.83 & \textbf{0.033} & 10 & $+7.5$ & 2.28 & 0.072 & 8 & $+25.2$ & 4.65 & \textbf{0.004} & 10 \\
\bottomrule
\end{tabular}
\end{table}

\clearpage

\subsection{Routing sweep}
\label{app:routing-sweep}

\begin{figure}[p]
    \centering
    \includegraphics[width=\linewidth]{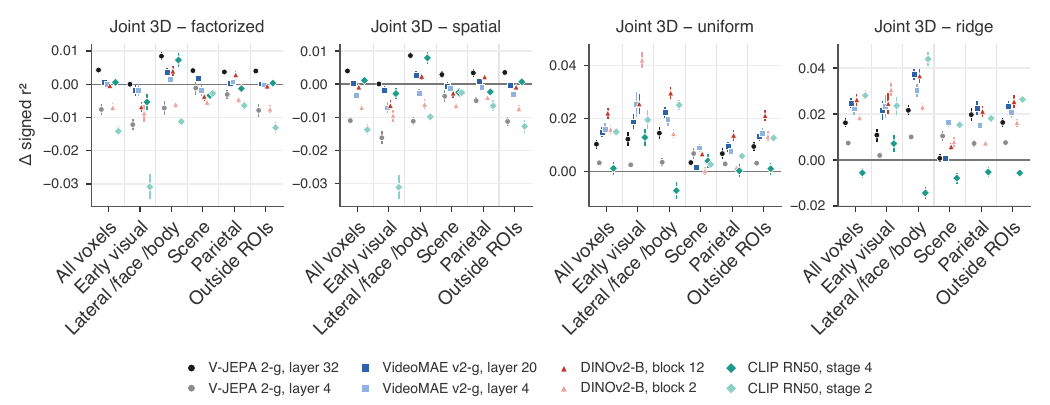}
    \caption{Joint model minus the factorized, spatial and mean-pool (``uniform'') models and ridge regression on mean-pooled features of the same layer, in region-mean signed $r^2$ on the 102 test clips, for every backbone and layer of the routing sweep (mean $\pm$ SEM over 10 subjects; voxels with noise ceiling $\geq$ 5\%).}
    \label{fig:app-sweep-deltas}
\end{figure}

\begin{figure}[p]
    \centering
    \includegraphics[width=\linewidth]{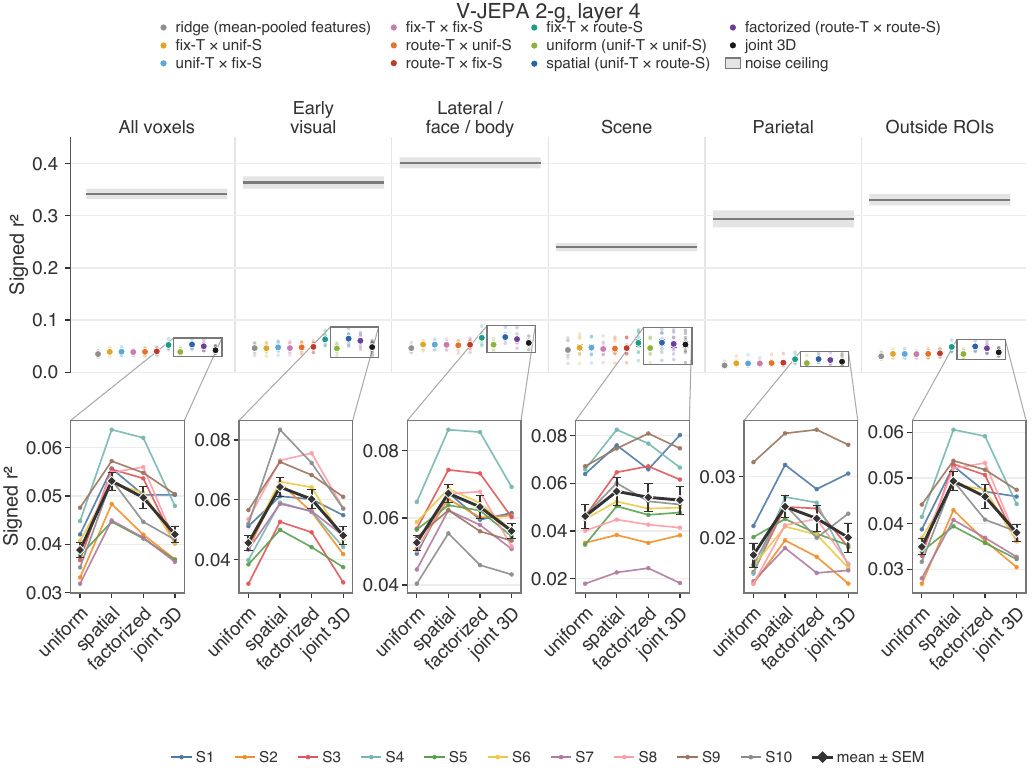}
    \caption{Fig.~\ref{fig:fig2}B,C for V-JEPA~2 ViT-g, layer 4 of 40: region-mean signed $r^2$ of every routing configuration, the joint model and ridge regression (top), and of the mean-pool, spatial, factorized and joint models in each subject (bottom).}
    \label{fig:app-sweep-vjepa-l4}
\end{figure}

\begin{figure}[p]
    \centering
    \includegraphics[width=\linewidth]{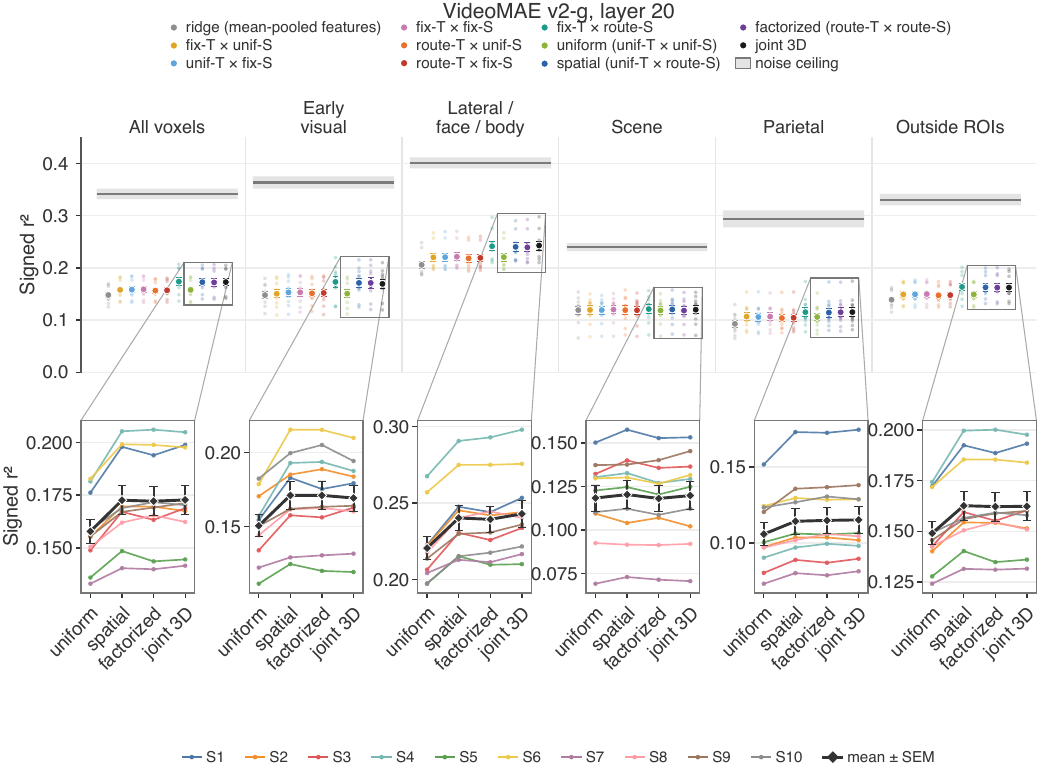}
    \caption{As Fig.~\ref{fig:app-sweep-vjepa-l4}, for VideoMAE~v2 ViT-g, layer 20 of 40.}
    \label{fig:app-sweep-vmae-l20}
\end{figure}

\begin{figure}[p]
    \centering
    \includegraphics[width=\linewidth]{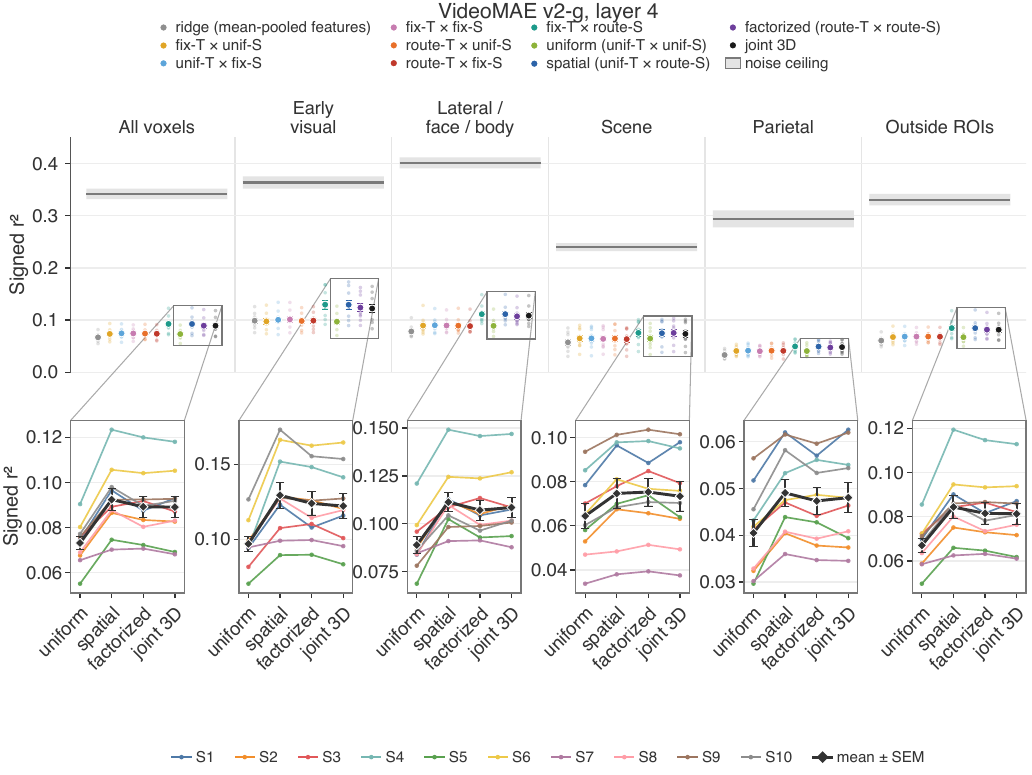}
    \caption{As Fig.~\ref{fig:app-sweep-vjepa-l4}, for VideoMAE~v2 ViT-g, layer 4 of 40.}
    \label{fig:app-sweep-vmae-l4}
\end{figure}

\begin{figure}[p]
    \centering
    \includegraphics[width=\linewidth]{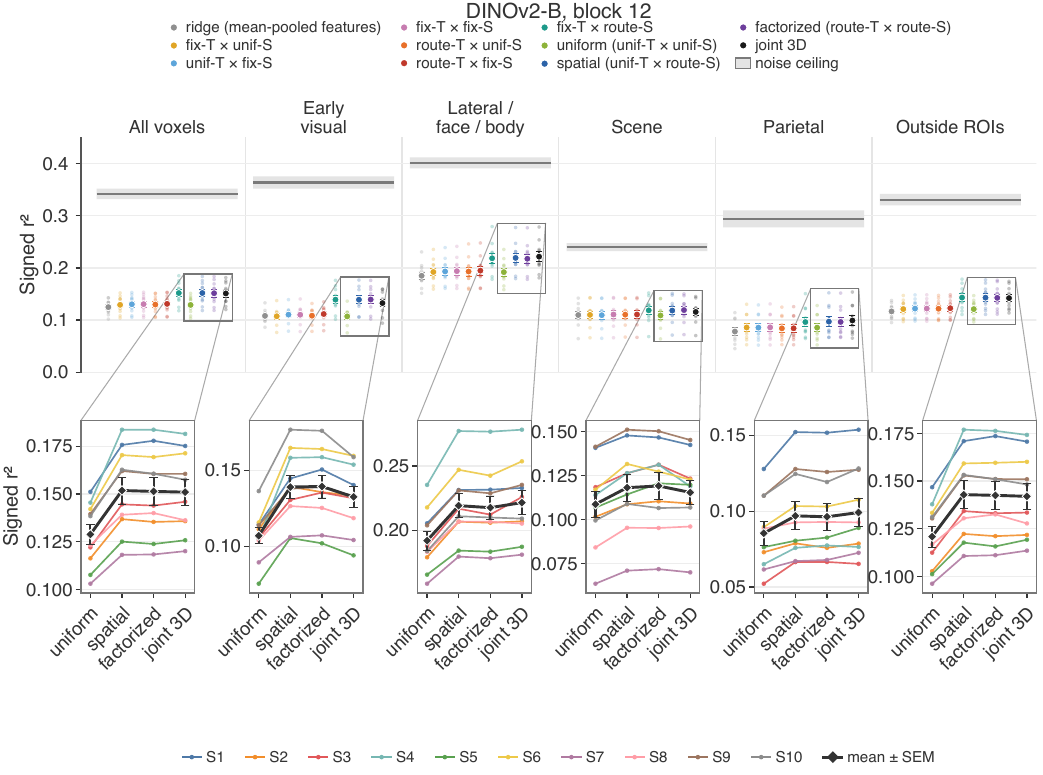}
    \caption{As Fig.~\ref{fig:app-sweep-vjepa-l4}, for DINOv2 ViT-B/14, block 12 of 12.}
    \label{fig:app-sweep-dinov2-b12}
\end{figure}

\begin{figure}[p]
    \centering
    \includegraphics[width=\linewidth]{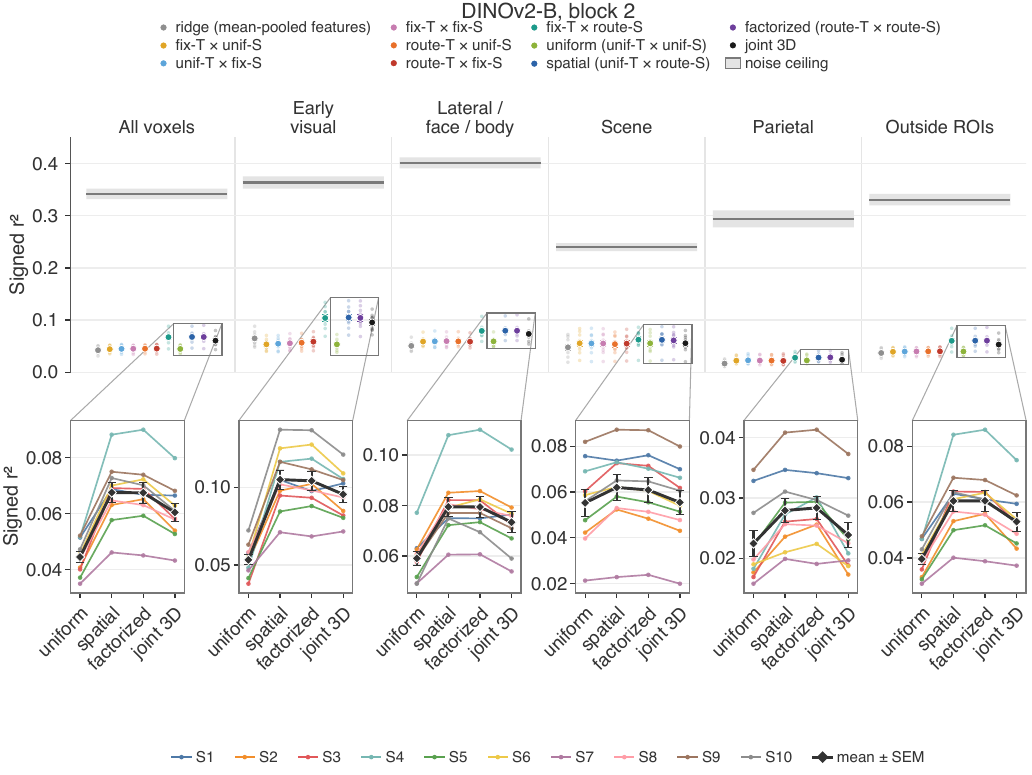}
    \caption{As Fig.~\ref{fig:app-sweep-vjepa-l4}, for DINOv2 ViT-B/14, block 2 of 12.}
    \label{fig:app-sweep-dinov2-b2}
\end{figure}

\begin{figure}[p]
    \centering
    \includegraphics[width=\linewidth]{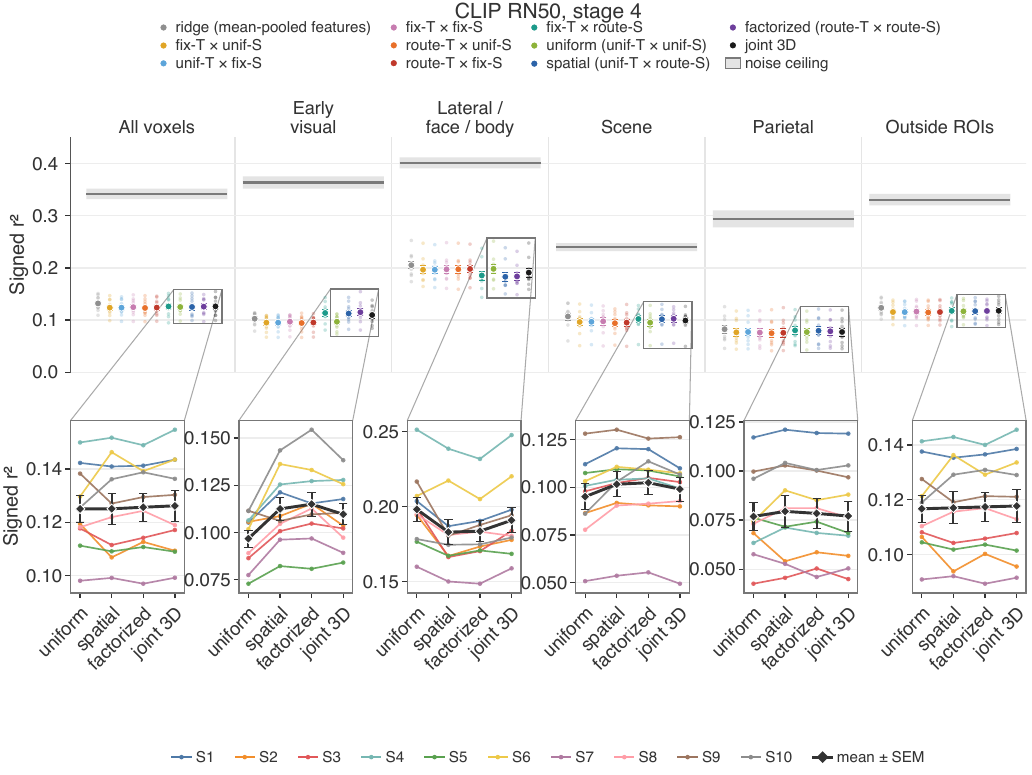}
    \caption{As Fig.~\ref{fig:app-sweep-vjepa-l4}, for CLIP RN50, stage 4 of 4.}
    \label{fig:app-sweep-clip-s4}
\end{figure}

\begin{figure}[p]
    \centering
    \includegraphics[width=\linewidth]{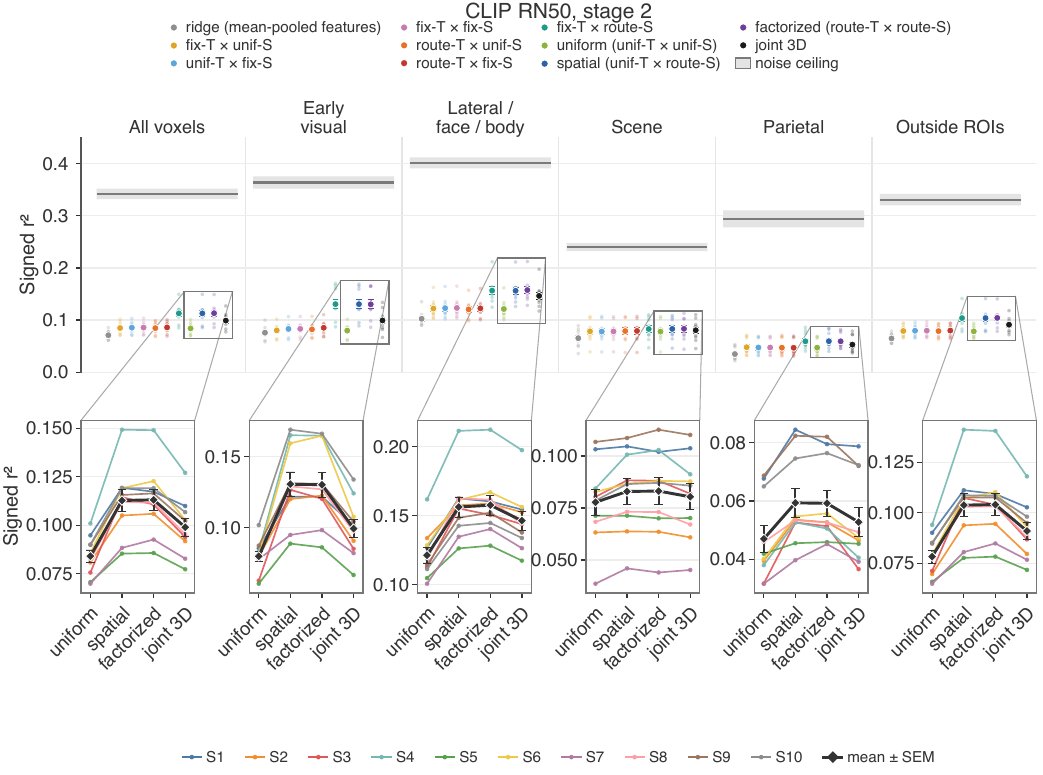}
    \caption{As Fig.~\ref{fig:app-sweep-vjepa-l4}, for CLIP RN50, stage 2 of 4.}
    \label{fig:app-sweep-clip-s2}
\end{figure}

Table~\ref{tab:routing-sweep} gives the region-mean signed $r^2$ of every routing configuration and of ridge regression for two layers of each of four backbones, all fit and evaluated as for V-JEPA~2 layer 32 (Section~\ref{sec:methods-fitting}). Figure~\ref{fig:app-sweep-deltas} summarizes the table as the difference between the joint model and the factorized, spatial and mean-pool models and ridge regression. Figures~\ref{fig:app-sweep-vjepa-l4}--\ref{fig:app-sweep-clip-s2} show every configuration other than V-JEPA~2 layer 32 in the format of Fig.~\ref{fig:fig2}B,C.

\begin{table}[h]
\centering
\scriptsize
\setlength{\tabcolsep}{3.5pt}
\caption{Region-mean signed $r^2$ on the 102 test clips (mean over 10 subjects; voxels with noise ceiling $\geq$ 5\%) for every routing configuration and backbone. The factorized models are labelled by their temporal factor (column groups) and spatial factor (U = uniform, F = fixed, R = routed), where a fixed factor is a learned, stimulus-independent distribution for each parcel. Temporal uniform with spatial uniform is the mean-pool model, temporal uniform with spatial routed is the spatial model, temporal routed with spatial uniform is the temporal model, and temporal routed with spatial routed is the factorized model. Ridge is ridge regression on the mean-pooled features of the same layer. The best model in each row is in bold.}
\label{tab:routing-sweep}
\begin{tabular}{lrrrrrrrrrrr}
\toprule
 & & \multicolumn{3}{c}{\textbf{Temporal uniform}} & \multicolumn{3}{c}{\textbf{Temporal fixed}} & \multicolumn{3}{c}{\textbf{Temporal routed}} & \\
\cmidrule(lr){3-5} \cmidrule(lr){6-8} \cmidrule(lr){9-11}
\textbf{Region} & \textbf{Ridge} & \textbf{U} & \textbf{F} & \textbf{R} & \textbf{U} & \textbf{F} & \textbf{R} & \textbf{U} & \textbf{F} & \textbf{R} & \textbf{Joint} \\
\midrule
\multicolumn{12}{l}{\textit{V-JEPA~2 ViT-g \citep{assran2025vjepa2selfsupervisedvideo}, layer 32 of 40}} \\
All voxels & .174 & .180 & .181 & .187 & .181 & .181 & .187 & .180 & .180 & .186 & \textbf{.191} \\
Early visual & .178 & .177 & .178 & .189 & .177 & .177 & \textbf{.190} & .178 & .179 & .189 & .189 \\
Lateral/face/body & .242 & .249 & .250 & .255 & .250 & .250 & .255 & .248 & .249 & .255 & \textbf{.263} \\
Scene & .118 & .115 & .117 & .116 & .117 & .117 & .116 & .114 & .114 & .115 & \textbf{.119} \\
Parietal & .109 & .122 & .122 & .125 & .122 & .122 & .126 & .121 & .122 & .125 & \textbf{.129} \\
Outside ROIs & .165 & .172 & .172 & .178 & .172 & .172 & .178 & .171 & .172 & .177 & \textbf{.181} \\
\midrule
\multicolumn{12}{l}{\textit{V-JEPA~2 ViT-g, layer 4 of 40}} \\
All voxels & .035 & .039 & .039 & \textbf{.053} & .039 & .039 & .052 & .039 & .040 & .050 & .042 \\
Early visual & .046 & .046 & .047 & \textbf{.064} & .046 & .046 & .063 & .048 & .049 & .060 & .048 \\
Lateral/face/body & .046 & .053 & .053 & \textbf{.067} & .053 & .052 & .066 & .052 & .053 & .063 & .056 \\
Scene & .043 & .046 & .047 & \textbf{.057} & .047 & .044 & .056 & .045 & .046 & .054 & .053 \\
Parietal & .013 & .017 & .017 & \textbf{.025} & .017 & .017 & \textbf{.025} & .018 & .018 & .023 & .020 \\
Outside ROIs & .030 & .035 & .035 & \textbf{.049} & .035 & .035 & .048 & .035 & .036 & .046 & .038 \\
\midrule
\multicolumn{12}{l}{\textit{VideoMAE~v2 ViT-g \citep{wang2023videomaev2}, layer 20 of 40}} \\
All voxels & .148 & .158 & .158 & .173 & .158 & .159 & \textbf{.174} & .156 & .157 & .172 & .173 \\
Early visual & .148 & .151 & .153 & .171 & .150 & .153 & \textbf{.173} & .151 & .152 & .171 & .169 \\
Lateral/face/body & .206 & .221 & .221 & .240 & .220 & .221 & .242 & .218 & .219 & .239 & \textbf{.243} \\
Scene & .119 & .118 & .119 & .120 & .119 & .120 & \textbf{.121} & .119 & .119 & .118 & .120 \\
Parietal & .093 & .106 & .106 & .114 & .107 & .107 & \textbf{.115} & .104 & .104 & \textbf{.115} & \textbf{.115} \\
Outside ROIs & .139 & .149 & .149 & .163 & .149 & .150 & \textbf{.164} & .147 & .148 & .162 & .162 \\
\midrule
\multicolumn{12}{l}{\textit{VideoMAE~v2 ViT-g, layer 4 of 40}} \\
All voxels & .067 & .073 & .074 & \textbf{.092} & .073 & .074 & \textbf{.092} & .074 & .074 & .089 & .089 \\
Early visual & .099 & .097 & .100 & \textbf{.129} & .097 & .101 & \textbf{.129} & .098 & .099 & .124 & .122 \\
Lateral/face/body & .078 & .089 & .090 & \textbf{.111} & .089 & .089 & \textbf{.111} & .089 & .088 & .107 & .108 \\
Scene & .057 & .064 & .065 & \textbf{.075} & .065 & .064 & \textbf{.075} & .064 & .063 & \textbf{.075} & .073 \\
Parietal & .033 & .040 & .041 & \textbf{.049} & .041 & .040 & \textbf{.049} & .041 & .041 & .047 & .048 \\
Outside ROIs & .061 & .067 & .068 & \textbf{.084} & .067 & .068 & \textbf{.084} & .068 & .068 & .082 & .081 \\
\midrule
\multicolumn{12}{l}{\textit{DINOv2 ViT-B/14 \citep{oquab2023dinov2}, block 12 of 12}} \\
All voxels & .125 & .129 & .130 & \textbf{.152} & .129 & .130 & \textbf{.152} & .130 & .131 & .151 & .151 \\
Early visual & .108 & .107 & .110 & \textbf{.139} & .107 & .110 & \textbf{.139} & .108 & .112 & \textbf{.139} & .132 \\
Lateral/face/body & .185 & .192 & .193 & .219 & .192 & .194 & .218 & .193 & .195 & .218 & \textbf{.222} \\
Scene & .110 & .109 & .109 & .118 & .110 & .111 & \textbf{.119} & .110 & .111 & \textbf{.119} & .115 \\
Parietal & .078 & .085 & .085 & .097 & .086 & .086 & .096 & .084 & .084 & .096 & \textbf{.099} \\
Outside ROIs & .117 & .121 & .122 & \textbf{.143} & .121 & .122 & \textbf{.143} & .122 & .123 & \textbf{.143} & .142 \\
\midrule
\multicolumn{12}{l}{\textit{DINOv2 ViT-B/14, block 2 of 12}} \\
All voxels & .042 & .045 & .045 & \textbf{.068} & .044 & .045 & .067 & .045 & .045 & .067 & .060 \\
Early visual & .065 & .053 & .055 & \textbf{.105} & .053 & .055 & .104 & .056 & .058 & .104 & .095 \\
Lateral/face/body & .050 & .059 & .059 & \textbf{.079} & .059 & .059 & \textbf{.079} & .059 & .059 & \textbf{.079} & .073 \\
Scene & .048 & .055 & .055 & \textbf{.062} & .055 & .055 & \textbf{.062} & .054 & .055 & .061 & .055 \\
Parietal & .016 & .022 & .023 & \textbf{.028} & .022 & .022 & \textbf{.028} & .022 & .022 & \textbf{.028} & .024 \\
Outside ROIs & .037 & .040 & .040 & \textbf{.060} & .039 & .040 & \textbf{.060} & .040 & .040 & \textbf{.060} & .053 \\
\midrule
\multicolumn{12}{l}{\textit{CLIP RN50 \citep{radford2021learning}, stage 4 of 4}} \\
All voxels & \textbf{.132} & .125 & .124 & .125 & .124 & .125 & .126 & .123 & .124 & .126 & .126 \\
Early visual & .102 & .097 & .095 & .112 & .095 & .097 & .113 & .094 & .095 & \textbf{.115} & .110 \\
Lateral/face/body & \textbf{.205} & .198 & .196 & .183 & .196 & .197 & .185 & .197 & .198 & .184 & .191 \\
Scene & \textbf{.107} & .095 & .097 & .102 & .096 & .097 & .102 & .094 & .095 & .102 & .099 \\
Parietal & \textbf{.082} & .077 & .077 & .079 & .076 & .076 & .080 & .075 & .075 & .078 & .077 \\
Outside ROIs & \textbf{.123} & .117 & .115 & .117 & .115 & .116 & .118 & .115 & .115 & .117 & .118 \\
\midrule
\multicolumn{12}{l}{\textit{CLIP RN50, stage 2 of 4}} \\
All voxels & .071 & .084 & .085 & \textbf{.113} & .085 & .086 & \textbf{.113} & .084 & .086 & \textbf{.113} & .099 \\
Early visual & .076 & .080 & .083 & \textbf{.130} & .080 & .083 & \textbf{.130} & .082 & .085 & \textbf{.130} & .099 \\
Lateral/face/body & .102 & .121 & .123 & .156 & .122 & .123 & .156 & .120 & .122 & \textbf{.157} & .146 \\
Scene & .065 & .078 & .078 & \textbf{.083} & .078 & .078 & \textbf{.083} & .079 & .079 & \textbf{.083} & .080 \\
Parietal & .035 & .047 & .047 & \textbf{.059} & .048 & .047 & \textbf{.059} & .047 & .047 & \textbf{.059} & .053 \\
Outside ROIs & .065 & .078 & .079 & \textbf{.104} & .079 & .080 & \textbf{.104} & .079 & .080 & \textbf{.104} & .091 \\
\bottomrule
\end{tabular}
\end{table}

\clearpage

\subsection{Additional examples of attention maps}
\label{app:examples}

\begin{figure}[p]
    \centering
    \includegraphics[width=\linewidth]{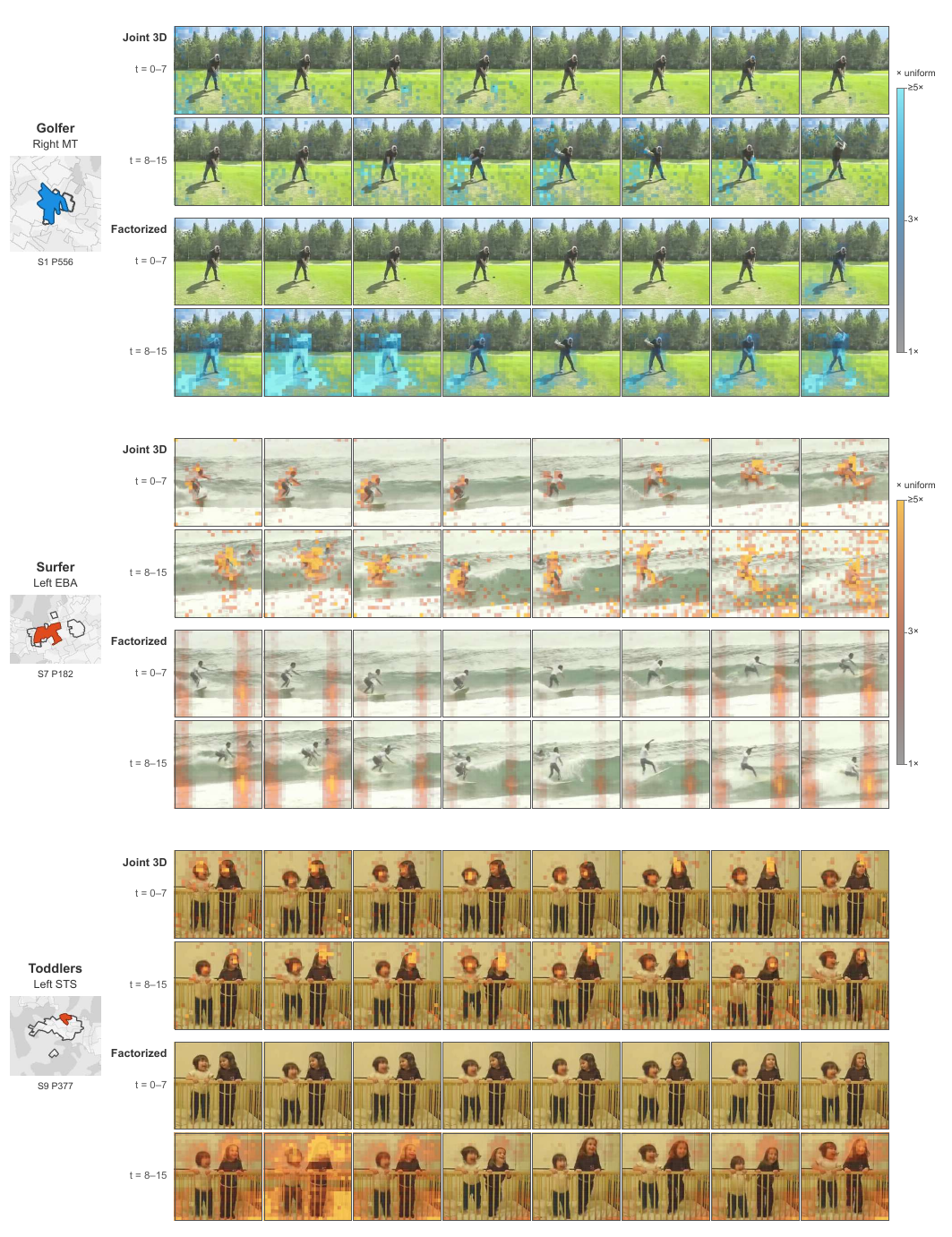}
    \caption{Further examples of attention maps in the format of Fig.~\ref{fig:fig3}: joint model (upper two rows of each block) and factorized model (lower two rows) over all 16 frames of each clip, in units of the uniform level $1/N$, from transparent at or below the uniform level to saturated at 5 times the uniform level.}
    \label{fig:app-examples-a}
\end{figure}

\begin{figure}[p]
    \centering
    \includegraphics[width=\linewidth]{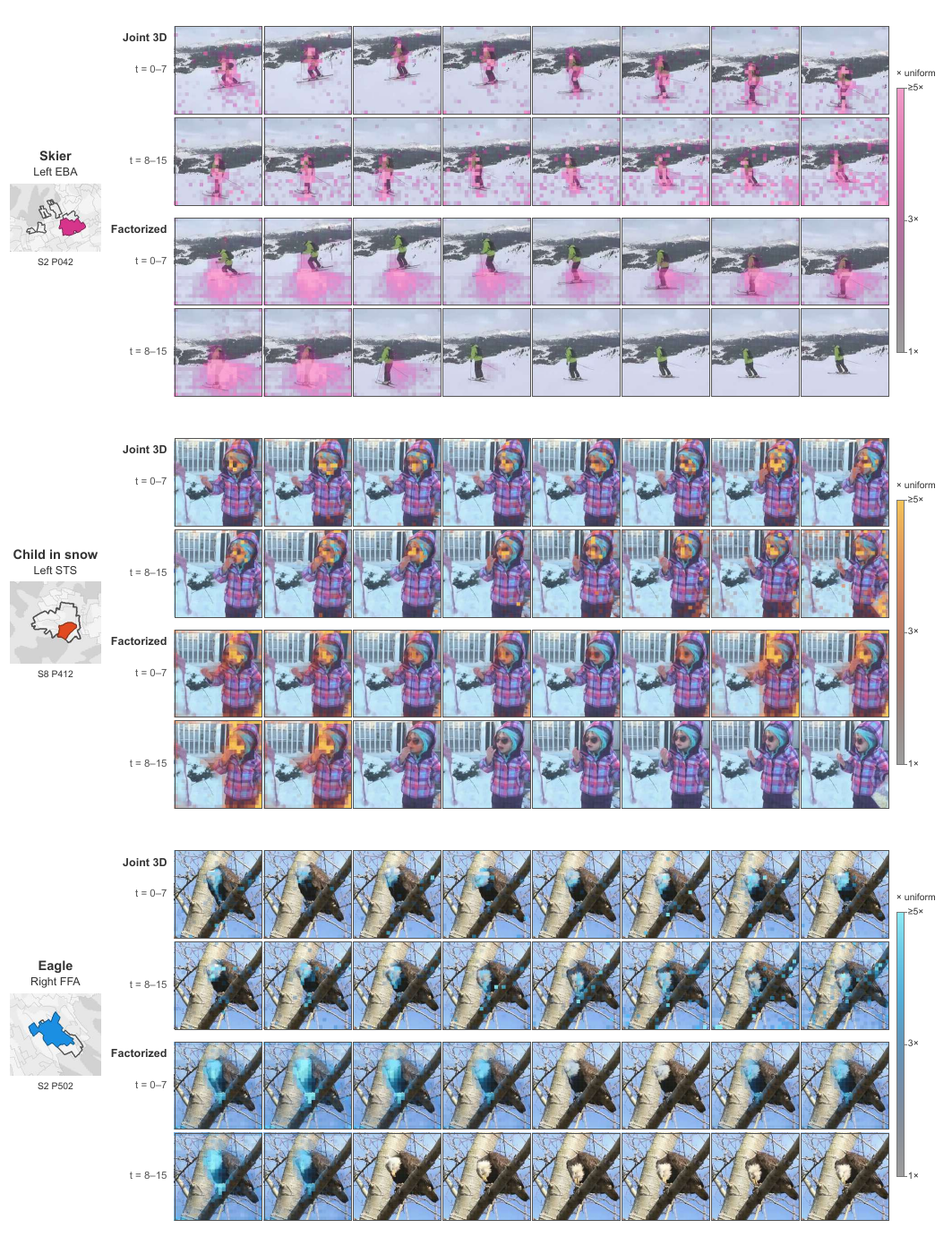}
    \caption{Further examples of attention maps, as in Fig.~\ref{fig:app-examples-a}.}
    \label{fig:app-examples-b}
\end{figure}

\begin{figure}[p]
    \centering
    \includegraphics[width=\linewidth]{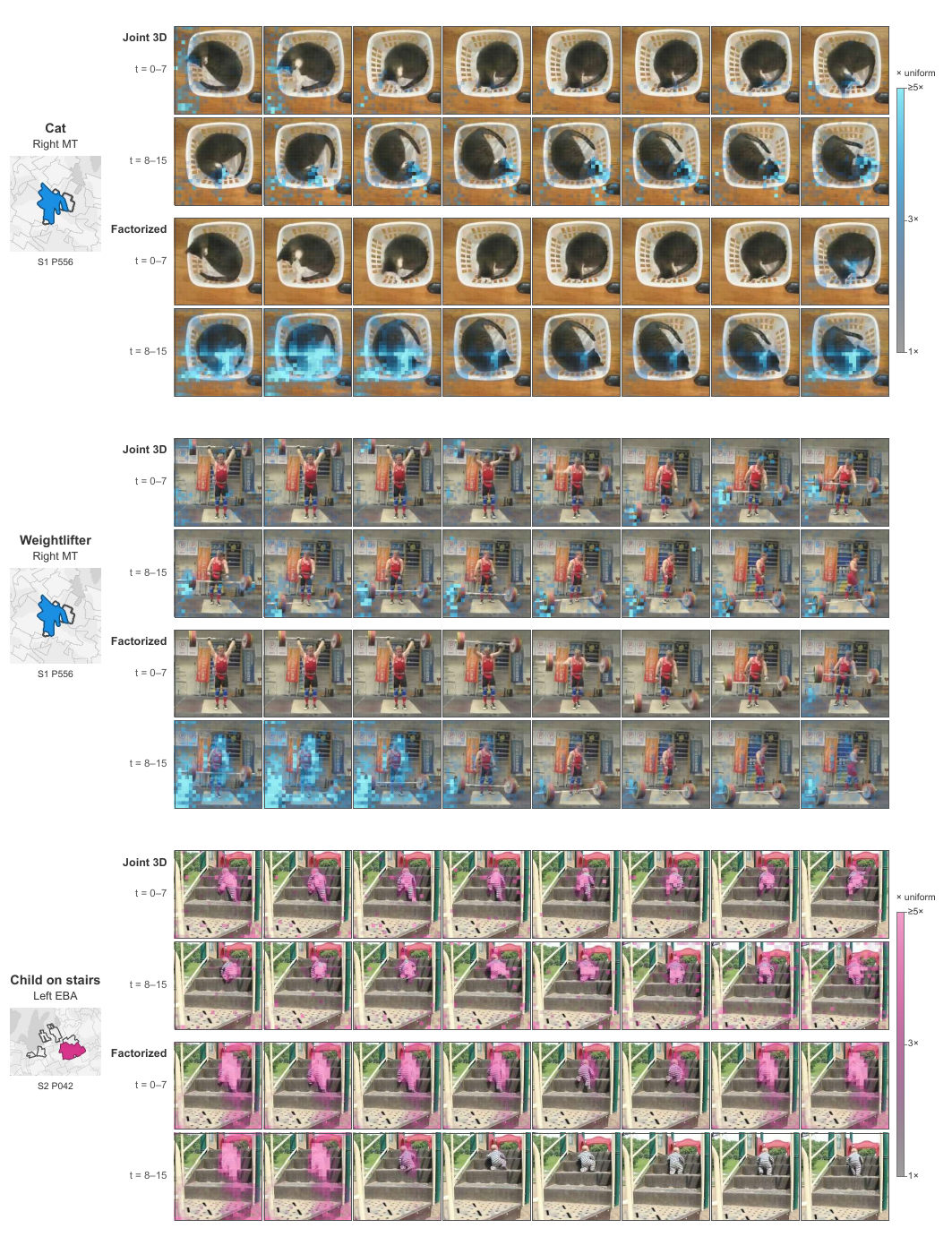}
    \caption{Further examples of attention maps, as in Fig.~\ref{fig:app-examples-a}.}
    \label{fig:app-examples-c}
\end{figure}

Figures~\ref{fig:app-examples-a}--\ref{fig:app-examples-c} show nine further test clips in the format of Fig.~\ref{fig:fig3}, for parcels in right MT (golfer, cat and weightlifter), left EBA (surfer, skier and child on stairs), left STS (toddlers and child in snow) and right FFA (eagle). We chose the clips by eye as examples of moving people and animals; they illustrate the attention maps and are not a quantitative test. For the golfer, the surfer, the skier, the toddlers and the child on stairs, the attention of the joint model stays on the moving person across frames, and for the cat it moves with the head and paw as the cat turns in its basket. The factorized model instead concentrates its attention on a few frames and, within them, on fixed parts of the frame that often include the background, such as the basket. For the weightlifter, the joint model attends mostly to the barbell as it is dropped and rolls, rather than to the lifter, whereas the factorized model spreads its attention over a large region on the left of the frame. In the eagle and child-in-snow clips, where the attended head moves little, the two models attend to similar locations.

\clearpage

\subsection{Category selectivity and localization from attention maps}
\label{app:catsel}

All analyses of category selectivity use the attention maps of the joint model (V-JEPA~2 ViT-g, layer 32) on the 102 test clips, with one model per subject trained as described in Section~\ref{sec:methods-fitting}. We assign a parcel to a functional ROI if at least 20\% of its voxels fall in that ROI. Because each parcel has a single attention map, the whole parcel is used, so a parcel can belong to more than one ROI and may also contain voxels outside the ROI.

\textbf{Category masks.} For each test clip we take 16 frames evenly spaced through the clip, one for each temporal position of the token grid, and segment them with SAM~3 \citep{carion2025sam3}, which returns all instances matching a text prompt; we keep detections with a confidence of at least 0.35. We use 28 prompts and merge them into mutually exclusive categories: face (face), body (person, hand), animal (animal, animal face), food (food), object (car, vehicle, ball, toy, musical instrument, tool, utensil, phone, book, furniture, bag, text, sign) and scene (building, tree, sky, grass, road, water, mountain, wall, floor). A pixel claimed by several categories is assigned to the first in this order. The order runs from the most to the least specific category, so that a face is not also counted as part of the person, and a hand holding a phone is not counted as an object. Pixels matched by no prompt are unlabeled. The masks are computed at $64 \times 64$ pixels per frame and area-averaged onto the $24 \times 24$ token grid, which gives the fraction $M_c(t,h,w) \in [0,1]$ of every token covered by category $c$. Because the categories are exclusive, these fractions sum to 1 over the categories and unlabeled. The unlabeled pixels are mostly scene surfaces that the scene prompts miss (for example snow, dirt paths, plain walls and courts), so when we score the known functional ROIs we merge scene and unlabeled into a single place category.

\textbf{Category selectivity of known functional ROIs.} We measure where a parcel reads in a clip by the 92 tokens (the top 1\% of the $16 \times 24 \times 24$ grid) that it attends to most, the same fraction as the 2,000 most-attended pixels of each image used by \citet{adeli2026transformerbrainencodersexplain}. Restricting to the most-attended tokens measures where the parcel reads, rather than the diffuse tail of its attention that is spread over the whole clip. The IoU of category $c$ is the intersection over union of these tokens with the binary mask of $c$ (tokens at least half covered), as in analyses of spatial selectivity in fMRI-trained networks \citep{sarch2023brain}. We average it over the clips in which $c$ covers between 1\% and 90\% of the clip: when a category covers very few tokens, its IoU is determined by those few tokens and dominated by noise, and when it fills nearly the whole clip, any 92 tokens overlap it, so the IoU carries no information about where the parcel attends. The raw IoU mostly reflects the size of each category (Fig.~\ref{fig:fig4}b). A place mask covers about 6,500 tokens, so 92 tokens can reach an IoU of only about 0.014 even for a parcel that attends only to places, whereas a face mask covers about 550 tokens and is hard to hit. We therefore z-score the IoU across parcels: the score of a functional ROI for category $c$ is the mean over its parcels minus the mean over all parcels, divided by the standard deviation across parcels. Because every parcel sees the same clips and masks, the size of each category is the same for all parcels and cancels, and the score asks whether the ROI puts more of its attention on the category than a typical parcel does. We report the raw and z-scored IoU for the eight category-selective ROIs (FFA, OFA, STS, EBA, PPA, RSC, TOS and LOC), and the raw IoU of all parcels for reference, for five categories (face, body, animal, object and place) as the mean $\pm$ SEM over subjects (Fig.~\ref{fig:fig4}b,c).

\textbf{Localizing category-selective regions from attention.} Functional localizers contrast the responses of each voxel to different stimulus categories, for example faces and objects. We apply the same logic to the attention maps: for each parcel we measure how much attention it puts on each category in each clip, form the localizer contrasts from these measures, and ask which parcels show the contrast consistently across clips.

\emph{Contrasts.} In each clip, we compute the IoU of each parcel's 92 most-attended tokens with the mask of each category, as above, and leave it undefined for a clip in which the category covers less than 1\% or more than 90\% of the clip. Here the scene category is the scene mask alone, without the unlabeled pixels. Within each subject, we z-score the IoU of each category across all clips and parcels where it is defined. The z-scoring puts the categories on a common scale, so that a contrast is not dominated by the category whose IoU is largest or varies most, which depends on its size. It also sets 0 to the average parcel on the average clip, so that a contrast measures a parcel's preference relative to other parcels rather than a preference that all parcels share (for example, if faces draw attention from every parcel). From the z-scored IoU we form, in each clip, contrasts that mirror the localizers used to define the functional ROIs: face minus object for FFA, OFA and STS, and body minus object for EBA. For PPA, RSC and TOS we contrast scene with the mean of face, body, animal and object. The scene localizer compares scenes with isolated objects, but in natural video every foreground item appears within a scene, so the analogous contrast is between scene and every kind of labeled foreground, not objects alone. A contrast is undefined in a clip if any of its terms is undefined; for the scene contrast, the mean is taken over the foreground categories that are defined. For each parcel and contrast, a one-sample $t$-test across clips then gives a $t$ value, treating clips as independent replicates, as trials are in a localizer. Parcels with fewer than 5 clips in which the contrast is defined are left unscored (Fig.~\ref{fig:fig4}a).

\emph{Scoring.} To compare these attention-derived maps with the functional ROIs, we pool the parcels of the ROIs that each localizer defines (face: FFA, OFA and STS; body: EBA; scene: PPA, RSC and TOS), with parcels assigned to ROIs as above. Let $K$ be the number of these ROI parcels. We take the $K$ scored parcels with the highest $t$ and measure the overlap as the fraction of them that belong to the ROIs (Fig.~\ref{fig:fig4}a shows the $t$ map, the top-$K$ parcels and the ROIs of each contrast for one subject, S8). Setting the size of the derived region equal to the size of the ROI avoids choosing a $t$ threshold and makes the overlap equal to both the precision and the recall of the derived region. If the ranking were unrelated to ROI membership, the number of ROI parcels among the top $K$ would follow a hypergeometric distribution, with an expected overlap equal to the fraction of scored parcels that belong to the ROIs (0.02--0.10). We therefore test the overlap in each subject with a one-sided hypergeometric test (uncorrected), and across subjects with a two-sided one-sample $t$-test of the overlap minus chance ($df = 9$), with $p$-values corrected over the three localizers with the Benjamini--Hochberg false discovery rate (FDR) procedure \citep{benjamini1995controlling}. As a threshold-free complement, we report the area under the ROC curve (AUC) of $t$ for ROI versus non-ROI scored parcels, which measures the separation of the two groups over the whole ranking.

\textbf{Detailed results.} The raw IoU mostly reflects the size of each category: faces, which are small, have the lowest IoU in most ROIs and in the average parcel, although the face IoU of FFA (0.0107) is well above that of all parcels (0.0063; Fig.~\ref{fig:fig4}b). After z-scoring (Fig.~\ref{fig:fig4}c), FFA, OFA and EBA put more attention than the average parcel on faces, bodies and animals ($z \approx 0.5$--$1.3$) and less on places; because these categories usually appear together in the clips, an area that attends to one of them also tends to attend to the others. PPA and RSC show the reverse pattern (place $z = 0.84$ and $0.75$, all foreground categories negative). LOC is high for faces, bodies and animals ($z = 1.15$--$1.25$), lower for objects ($z = 0.34$) and strongly negative for places ($z = -1.24$), consistent with a preference for foreground content in general. In STS and TOS, all categories are close to the average parcel ($|z| \leq 0.28$). In the localization, the overlap is significant in 5, 3 and 8 of the 10 individual subjects for the face, body and scene localizers (one-sided hypergeometric test, $p < 0.05$, uncorrected), and ROI parcels rank above other parcels across the whole ranking, not only at the top (AUC 0.61--0.77; Table~\ref{tab:iou-localization}). Fig.~\ref{fig:fig4}a shows the subject with the highest mean overlap across the three localizers (S8; face 0.41, body 0.08, scene 0.43).

\subsection{Spatial-prior control for the localization}
\label{app:simple-metrics}

The IoU of a parcel with a category can differ between parcels even if no parcel's attention follows the content of the clips, because each parcel's attention has a spatial prior and categories occupy typical positions in the frame. The IoU is computed from the most-attended tokens, and which tokens are most attended depends on the prior. Z-scoring does not remove this confound: unlike the size of a category, which is the same for all parcels and cancels, the prior differs between parcels. To measure this contribution, we repeat the localization with each clip's attention scored against the category masks of a different test clip. The pairing is a single random permutation of the 102 test clips in which no clip is paired with itself, drawn once and used for every parcel and subject. The mismatched masks keep each parcel's spatial prior and the typical layout of the categories, but remove any correspondence between the attention and the content of the clip. Table~\ref{tab:iou-localization} compares the localization with the correct and the mismatched masks; the two are compared with a two-sided paired $t$-test across subjects ($df = 9$), FDR-corrected over the three localizers.

For faces, the mismatched masks give an overlap at chance (0.10, chance 0.09; $t(9) = 0.25$, $p_{\mathrm{FDR}} = 0.81$), and the correct masks give a significantly larger overlap ($t(9) = 3.13$, $p_{\mathrm{FDR}} = 0.036$), so the localization of the face-selective ROIs depends on the content of each clip. For bodies, the mismatched-mask overlap (0.08) is not significantly above chance ($t(9) = 2.00$, $p_{\mathrm{FDR}} = 0.11$), but the correct masks do not exceed it significantly either ($t(9) = 1.70$, $p_{\mathrm{FDR}} = 0.18$). For scenes, the mismatched masks give nearly the same overlap as the correct masks (0.26 vs.\ 0.28; $t(9) = 1.11$, $p_{\mathrm{FDR}} = 0.30$) and are themselves well above chance ($t(9) = 3.74$, $p_{\mathrm{FDR}} = 0.014$), so the localization of the scene-selective ROIs mostly reflects the parcels' spatial priors rather than attention to the scenery in each clip.

\begin{table}[h]
\centering
\scriptsize
\setlength{\tabcolsep}{3pt}
\caption{Localization of category-selective ROIs from z-scored IoU contrasts with the correct masks and with the masks of a different clip (joint model, 10 subjects, 102 test clips; face: FFA, OFA and STS; body: EBA; scene: PPA, RSC and TOS). Overlap: fraction of the top-$K$ parcels that belong to the ROIs, mean (SD) over subjects; chance is the fraction of scored parcels in the ROIs. $t(9)$ and $p_{\mathrm{FDR}}$: two-sided one-sample $t$-test of overlap minus chance across subjects, Benjamini--Hochberg corrected over the three localizers within each condition. Subjects: number with a one-sided hypergeometric $p < 0.05$ (uncorrected). Correct $-$ mismatched: two-sided paired $t$-test of the overlaps across subjects, FDR-corrected over the three localizers.}
\label{tab:iou-localization}
\begin{tabular}{lrrrrrrrrrrrrr}
\toprule
 & & \multicolumn{5}{c}{\textbf{Correct masks}} & \multicolumn{4}{c}{\textbf{Mismatched masks}} & \multicolumn{3}{c}{\textbf{Correct $-$ mismatched}} \\
\cmidrule(lr){3-7} \cmidrule(lr){8-11} \cmidrule(lr){12-14}
\textbf{Localizer} & \textbf{Chance} & \textbf{Overlap} & \textbf{\boldmath$t(9)$} & \textbf{\boldmath$p_{\mathrm{FDR}}$} & \textbf{AUC} & \textbf{Subjects} & \textbf{Overlap} & \textbf{\boldmath$t(9)$} & \textbf{\boldmath$p_{\mathrm{FDR}}$} & \textbf{Subjects} & \textbf{\boldmath$t(9)$} & \textbf{\boldmath$p$} & \textbf{\boldmath$p_{\mathrm{FDR}}$} \\
\midrule
Face & 0.09 & 0.19 (0.11) & 2.81 & 0.021 & 0.61 & 5/10 & 0.10 (0.05) & 0.25 & 0.81 & 0/10 & 3.13 & 0.012 & 0.036 \\
Body & 0.02 & 0.14 (0.09) & 4.02 & 0.005 & 0.76 & 3/10 & 0.08 (0.09) & 2.00 & 0.11 & 3/10 & 1.70 & 0.12 & 0.18 \\
Scene & 0.10 & 0.28 (0.12) & 4.85 & 0.003 & 0.77 & 8/10 & 0.26 (0.14) & 3.74 & 0.014 & 8/10 & 1.11 & 0.30 & 0.30 \\
\bottomrule
\end{tabular}
\end{table}


\end{document}